# Enhanced Sampling of Protein Conformational Changes via True Reaction Coordinates from Energy Relaxation

Huiyu Li and Ao Ma*

Center for Bioinformatics and Quantitative Biology
Richard and Loan Hill Department of Biomedical Engineering
The University of Illinois Chicago
851 South Morgan Street
Chicago, IL 60607

*correspondence should be addressed to: Ao Ma
Email: aoma@uic.edu
Tel: (312) 996-7225

**Abstract**

The bottleneck in enhanced sampling lies in finding collective variables (CVs) that can effectively accelerate protein conformational changes. True reaction coordinates (tRCs) that can predict the committor are considered the optimal CVs, but identifying them requires unbiased natural reactive trajectories, which, paradoxically, depend on effective enhanced sampling. Using the generalized work functional method, we found that tRCs control both conformational changes and energy relaxation, enabling us to compute tRCs from energy relaxation simulations. Applying bias to tRCs accelerated conformational changes and ligand dissociation in HIV-1 protease and the PDZ2 domain by $10^5$ to $10^{15}$ -fold. The resulting trajectories follow natural transition pathways, enabling efficient generation of natural reactive trajectories. In contrast, biased trajectories from empirical CVs often display non-physical features. Furthermore, by computing tRCs from a single protein structure, our method enables predictive sampling of conformational changes. These findings significantly broaden the range of protein functional processes accessible to molecular dynamics simulations.

**Introduction**

The main goal of molecular biophysics is to understand how proteins function. Protein behavior is governed by a rugged energy landscape featuring numerous valleys separated by barriers [1]. The valleys correspond to functionally important metastable conformations, with the deepest valley being the native structure. Transitions between conformations are critical for protein function [1-3], such as enzymatic reactions, allostery, substrate binding, and protein-protein interaction.

With the development of AlphaFold [4,5], the long-standing structure prediction problem has been solved, making the native structures of proteins readily available. The major challenge now is to identify the other functionally important conformations and understand the transition dynamics between them. Molecular dynamics (MD) simulation is a pivotal tool, as it can provide full atomic details—an advantage unmatched by experimental techniques. However, due to the large gap between the time scales of MD simulation (approaching microseconds) and functional processes (milliseconds to hours), MD simulation of functional processes has been infeasible except in a few special cases [6,7].

To overcome this time scale challenge, intensive efforts have been focused on developing methods to enhance the sampling of functionally important regions of the conformational space, including two branches. One branch focuses on sampling important metastable conformations without addressing the transition dynamics between them [8,9]. The other branch focuses on sampling the transition dynamics between different metastable conformations [10,11].

Enhanced sampling of important conformations consists of two components: accelerating conformational changes to effectively explore the conformational space and reweighting to extract the correct thermodynamics. The former is the bottleneck and the focus of the current work. Many methods, such as umbrella sampling, adaptive biasing force, and metadynamics [12-15], use bias potentials on user-selected collective variables (CVs) to accelerate conformational changes. Their efficacy hinges on finding suitable CVs; without them, they provide no more acceleration than standard MD simulations [8]. Traditionally, CVs were chosen by user intuition, involving geometric parameters (e.g., radius of gyration), principal components, evolutionary correlations, and RMSD from reference structures [8,16,17]. In recent years, it has been increasingly clear that intuition is inadequate for identifying the suitable CVs [8,18]. Consequently, significant effort has been made to develop systematic methods for CV identification, yielding many innovative methods [19]. The predominant strategy was to use machine learning to extract slow modes from simulation data [18,20-22]. Despite these efforts, finding CVs that can effectively accelerate protein conformational changes remains a formidable challenge.

Among the methods for sampling transition dynamics, the transition path sampling (TPS) method is an important example [10,11]. It offers unique advantages over methods like milestoning, Markov state modeling, and string method [23-25] because it generates natural reactive trajectories (NRTs)—unbiased MD trajectories that connect the reactant and product basins without any assumption or approximation, providing the full details of transition dynamics that mirror the physical reality. NRTs also achieve optimal efficiency by covering the entire transition period while skipping the prolonged waiting period in the reactant basin. The waiting period is orders of magnitude longer than the transition period, causing MD simulations to be trapped in the reactant basin without reaching the transition region, thus preventing simulation of functional processes. TPS generates NRTs by initiating MD trajectories from conformations on existing NRTs close to the transition state (TS). However, finding an initial NRT or TS conformations poses a formidable challenge, and TPS does not provide an effective solution for this, hindering applications of TPS to complex systems.

The solutions to the bottleneck challenges in both branches of enhanced sampling are provided by true reaction coordinates (tRCs), the few essential protein coordinates that fully determine the committor of any system conformation [10,26,27]. The committor ($p_B$) is the probability that a trajectory initiated from a given system conformation, with momenta drawn from Boltzmann distribution, will reach the product state before the reactant state [10,26,28,29]. It precisely tracks the progression of a conformational change, marking the key states: $p_B = 0$ for the reactant, $p_B = 1$ for the product, and $p_B = 0.5$ for the transition state (TS). tRCs are coordinates that can accurately predict the committor for any conformation, rendering all other system coordinates irrelevant.

tRCs are widely regarded as the optimal CVs for accelerating conformational changes [8,9], as they are expected to not only provide efficient acceleration but also generate accelerated trajectories that follow the natural transition pathway. This is due to their role in energy activation—the critical step in protein conformational change where rare fluctuations channel energy into tRCs to propel the system over the activation barrier (i.e. the TS) to the product basin [30,31]. Applying bias potentials on tRCs maximizes energy transfer into tRCs, leading to highly accelerated barrier crossing. In contrast, if the CVs significantly deviate from the tRCs, the bias potential will miss the actual activation barrier, resulting in the infamous "hidden barrier" [32,33] that prevents effective sampling.

Furthermore, tRCs provide an effective solution to the bottleneck problem in TPS. As we showed in Fig. 7a of ref. [34], bias applied to tRCs generates trajectories that pass through the full range of intermediate committor values $p_B \in [0.1, 0.9]$, which marks the transition period. This provides an efficient means to obtain TS conformations, which are otherwise difficult to attain, thereby addressing the key limitation of TPS.

Given the importance of tRCs, identifying them in complex biomolecules has been a central challenge in chemical physics and molecular biophysics since the pioneering work by Du et. al. and Chandler and colleagues in the late 1990s [10,19,26,27,35-39]. In recent years, we have made important progress by developing rigorous, physics-based methods—energy flow theory and the generalized work functional (GWF) method [31,40-43]. Using these methods, we have identified the tRCs for the flap opening process of HIV-1 protease (HIV-PR) in implicit solvent [34], marking the first successful identification of tRCs in proteins.

Previously, tRCs were identified only from NRTs, which require effective enhanced sampling. As a result, tRCs were not useful in solving the bottlenecks in enhanced sampling. In this work, we showed that tRCs control both conformational changes and energy relaxation, allowing us to compute tRCs from the energy relaxation of a single protein structure. Applying bias potentials to the tRCs efficiently accelerates protein conformational changes. The resulting trajectories, termed RC-uncovered trajectories, follow natural transition pathways and pass through TS conformations, thereby enabling the generation of NRTs using TPS. Therefore, our findings provide an effective solution to the bottlenecks in both accelerating protein conformational changes and harvesting NRTs. Furthermore, our method computes tRCs from a single protein structure, giving it predictive power.

Applying our method to HIV-PR in explicit solvent, we obtained RC-uncovered trajectories that achieve flap opening and ligand unbinding, a process with an experimental lifetime of $8.9 \times 10^5$ s, in 200 ps. Furthermore, we generated NRTs by applying the shooting move of TPS to these trajectories, confirming that they follow natural transition pathways. In contrast, biased trajectories for this process using a commonly adopted CV follow a non-physical transition pathway, reinforcing the widely accepted assumption that tRCs are optimal CVs for enhanced sampling. To test our method's predictive power, we applied it to PDZ domain allostery, a long-standing puzzle for over 20 years [44]. From the NRTs of ligand dissociation from PDZ2 domain, we observed previously unknown large-scale transient conformational changes at PDZ allosteric sites, offering an intuitive mechanism for PDZ allostery—effectors affect ligand binding to PDZ domains by interfering with these fleeting conformational changes.

**Results**

We use two methods to identify tRCs. First, potential energy flows (PEFs) through individual coordinates measure the importance of each coordinate in driving protein conformational changes, with the highest PEFs indicating the critical coordinates. Second, the generalized work functional (GWF) generates an orthonormal coordinate system, called singular coordinates (SCs), that disentangles tRCs from non-RCs by maximizing the PEFs through individual coordinates. Consequently, tRCs are identified as the SCs with the highest PEFs [34,40,43].

**Potential energy flow**. The motion of a coordinate is governed by its equation of motion (EoM). For a coordinate $q_i$, its EoM is generated by the mechanical work done on $q_i$, given by: $dW_i = -\frac{\partial U(\boldsymbol{q})}{\partial q_i} dq_i$ [40]. The EoM of $q_i$ is: $\dot{p}_i = -\frac{\partial H}{\partial q_i} = -\frac{\partial K}{\partial q_i} - \frac{\partial U}{\partial q_i}$, where $p_i$ is the conjugate momentum of $q_i$, $K(\boldsymbol{q}, \dot{\boldsymbol{q}})$ is the kinetic energy, and $U(\boldsymbol{q})$ is the potential energy of the system. Moving $\frac{\partial K}{\partial q_i}$ to the left-hand side results in:

$$\dot{p}_i + \frac{\partial K}{\partial q_i} = -\frac{\partial U(\boldsymbol{q})}{\partial q_i} = \frac{dW_i}{dq_i} \quad (1),$$

The left-hand side of Eq. (1) represents the total change in the motion of $q_i$, meaning that $dW_i$ is the energy cost of the motion of $q_i$, which we refer to as the PEF through $q_i$. The PEF of $q_i$ during a finite period is given by the integration of $dW_i$: $\Delta W_i(t_1, t_2) = \int_{t_1}^{t_2} dW_i = -\int_{q_i(t_1)}^{q_i(t_2)} \frac{\partial U(\boldsymbol{q})}{\partial q_i} dq_i$.

Intuitively, coordinates with a higher energy cost require more effort from the system and play a more significant role in dynamic processes. In protein conformational changes, which are activated processes, tRCs need to overcome the activation barrier, incurring the highest energy cost for their motion. Therefore, tRCs correspond to coordinates with the highest PEFs.

**Generalized work functional**. The remaining challenge in identifying tRCs is the entanglement between tRCs and non-RCs in general-purpose coordinate systems (e.g., internal coordinates), where many coordinates have components in both tRC and non-RC directions. To cleanly distinguish tRCs from non-RCs, we need to transform internal coordinates into an "ideal" coordinate system that disentangles them. Since PEFs measure the importance of individual coordinates, this "ideal" coordinate system should maximize the differences in PEFs, thereby maximally separating the most important coordinates (tRCs) from the unimportant ones (non-RCs). This can be achieved by maximizing the PEFs through individual coordinates in an orthogonal coordinate system.

Finding the "ideal" coordinate system starts with understanding how PEFs transform between different coordinate systems. The transformation of $dW_i = F_i dq_i$ in coordinates $\boldsymbol{q}$ to $dW_\alpha = F_\alpha dr_\alpha$ in coordinates $\boldsymbol{r}$, related by $d\boldsymbol{r} = \boldsymbol{A} \cdot d\boldsymbol{q}$ and $d\boldsymbol{q} = \boldsymbol{A}^{-1} \cdot d\boldsymbol{r}$, is given by the chain rule:

$$dW_\alpha = -\frac{\partial U}{\partial r_\alpha} dr_\alpha = -\sum_{i,k=1}^{N} \frac{\partial U}{\partial q_i} \frac{\partial q_i}{\partial r_\alpha} \frac{\partial r_\alpha}{\partial q_k} dq_k = \sum_{i,k=1}^{N} A_{\alpha i}(F_i dq_k) A_{k\alpha}^{-1} \quad (2).$$

Here, $A_{\alpha i} = A_{i\alpha}^{-1} = \frac{\partial r_\alpha}{\partial q_i} = \frac{\partial q_i}{\partial r_\alpha}$ because $\boldsymbol{A}$ is an orthogonal matrix. Equation (2) introduces a conceptually new physical quantity $F_i dq_k$, which is not a mechanical work. Therefore, the coordinate transformation of $dW_i$ requires generalizing the concept of mechanical work to incorporate quantities like $F_i dq_k$.

The GWF generalizes the concept of mechanical work. Its differential in coordinates $q$ is defined as:

$$d\mathbb{W}_q = \mathbf{F} \otimes d\mathbf{q} \quad (3).$$

Here, $\otimes$ denotes tensor product, making $d\mathbb{W}_q$ an asymmetric tensor and $F_i dq_k$ its elements. The GWF in coordinates $\mathbf{r}$ and $\mathbf{q}$ are related by a similarity transformation:

$$d\mathbb{W}_r = \mathbf{A} \cdot d\mathbb{W}_q \cdot \mathbf{A}^{-1} \quad (4).$$

Therefore, the coordinate transformation of GWF does not introduce any new quantities, making it a self-contained fundamental concept, with mechanical work as its sub-concept. All the important mechanical quantities, such as $d\mathbf{W} = (dW_1, dW_2, \ldots, dW_N)$ and $dU$, are encompassed in GWF:

$$d\mathbf{W}_r = diag(d\mathbb{W}_r) = diag(\mathbf{A} \cdot d\mathbb{W}_q \cdot \mathbf{A}^{-1}) \quad (5)$$

$$dU = Tr(d\mathbb{W}_r) = Tr(d\mathbb{W}_q) \quad (6),$$

where $diag(\cdot)$ denotes diagonal vector and $Tr(\cdot)$ denotes trace. Equation (5) further shows that $d\mathbb{W}$ is the operator for transforming $d\mathbf{W}$ between different coordinate systems.

The singular value decomposition of GWF is:

$$d\mathbb{W} = \mathbf{U} \cdot \mathbf{\Lambda} \cdot \mathbf{V}^T \quad (7).$$

It decomposes $d\mathbb{W}$ into its optimal basis tensors $\mathbf{u}_i \otimes \mathbf{v}_i$ as: $d\mathbb{W} = \sum_{i=1}^N \lambda_i \mathbf{u}_i \otimes \mathbf{v}_i$. Here, $\mathbf{u}_i$ and $\mathbf{v}_i$ are the $i$-th column vectors of the left and right singular matrices $\mathbf{U}$ and $\mathbf{V}$, respectively; $\lambda_i$ is the $i$-th singular value. The optimality of $\mathbf{u}_i \otimes \mathbf{v}_i$ means that $\lambda_i \mathbf{u}_i \otimes \mathbf{v}_i$ represents the $i$-th largest contribution to $d\mathbb{W}$, and the sum $\sum_{i=1}^{m \ll N} \lambda_i \mathbf{u}_i \otimes \mathbf{v}_i$ provides the optimal $m$-dimensional reduced description. Since $d\mathbb{W} = \mathbf{F} \otimes d\mathbf{q}$, $\mathbf{u}_i$ and $\mathbf{v}_i$ are the optimal basis vectors for the force and displacement spaces, respectively.

The collection of all $\mathbf{u}_i$ forms an orthonormal coordinate system: $d\mathbf{u} = \mathbf{U}^T \cdot d\mathbf{q}$, which we refer to as the SCs. $F_i = -\frac{\partial U}{\partial s_i}$ represents the force with the $i$-th largest impact on the system's dynamics, and $dW_i = F_i ds_i = \lambda_i(\mathbf{u}_i \cdot \mathbf{v}_i)$ is the $i$-th highest PEF in the system. Consequently, $\sum_{i=1}^{m \ll N} dW_i$ provides the optimal $m$-dimensional reduced description of $dU$, the generating function of all equations of motion in the system. This is the condition tRCs would meet if they exist in a protein.

Therefore, the singular coordinates provide the "ideal" coordinate system; the leading singular coordinates are tRCs, as they represent the directions of forces with the highest impact on the system's dynamics.

**A hypothesis on activation and energy relaxation.** In ref. [34], we demonstrated that the leading SCs of the GWF calculated from an ensemble of NRTs for the flap-opening of HIV-PR in implicit solvent are tRCs, as they can determine the committor with high accuracy. This result confirms that the GWF method can effectively identify tRCs in proteins. In addition, applying bias potentials along the tRCs led to RC-uncovered trajectories that open the flaps within 4 ps—an acceleration of over $10^4$-fold compared to MD [34]. This result supports the long-standing hypothesis that tRCs are the optimal CVs for enhanced sampling of protein conformational changes.

In our application of the GWF method to energy relaxation in myoglobin [42], we observed a significant gap between the PEFs of the leading SCs and the others, resembling the PEF pattern of tRCs in ref [34]. (Fig. 2C in ref. [34]). Energy relaxation and activation are closely related, as a system needs to relax into the stable basin after crossing the activation barrier. Drawing on Onsager's regression hypothesis [45], we propose that the coordinates essential for energy relaxation (i.e. its leading SCs) are identical to those governing activation (i.e. the tRCs).

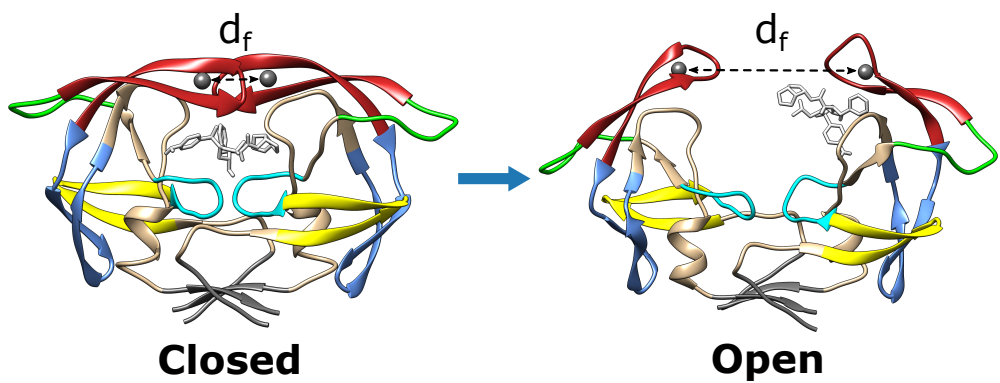

**Fig. 1: Representative structures of the closed and open states of DRV-bound HIV-PR.** The closed structure ($d_f = 1.03$ nm) is from crystal structure (Protein Data Bank: 1T3R), and the open structure ($d_f = 2.73$ nm) is from a natural trajectory of DRV dissociation. Different structural segments are indicated by different colors. The two spheres denote the center of mass of the $C_\alpha$ atoms of residues Met46 to Lys55 in the two monomers; the flap distance $d_f$ is the distance between them.

**Leading SCs of energy relaxation are the same as tRCs of conformational transition**. To determine if the leading SCs of HIV-PR energy relaxation are also the tRCs for its activation (e.g. flap opening; Fig. 1), we simulated energy relaxation of ligand-free HIV-PR in implicit solvent, the same system as in ref. [34]. To mimic energy relaxation after ligand binding or enzymatic reaction, we deposit excess kinetic energy into the active site of HIV-PR (details in the Methods) and generate an ensemble of $N$ MD trajectories, each one 5 ps in length [42]. We then compute GWF and potential energy flows averaging over this ensemble.

$$\langle \Delta \mathbb{W}(0 \to t) \rangle = \frac{1}{N} \sum_{\alpha=1}^{N} \Delta \mathbb{W}(0 \to t; \alpha) \quad (8)$$

$$\langle \Delta W_i(0 \to t) \rangle = \frac{1}{N} \sum_{\alpha=1}^{N} \Delta W_i(0 \to t; \alpha) \quad (9).$$

Here, $t = 0$ is the starting point of energy relaxation when the excess kinetic energy is deposited to the active site; $\Delta \mathbb{W}(0 \to t; \alpha) = \int_0^t d\mathbb{W}$ and $\Delta W_i(0 \to t; \alpha) = \int_0^t dW_i$ are GWF and mechanical work on $q_i$ along trajectory $\alpha$, respectively.

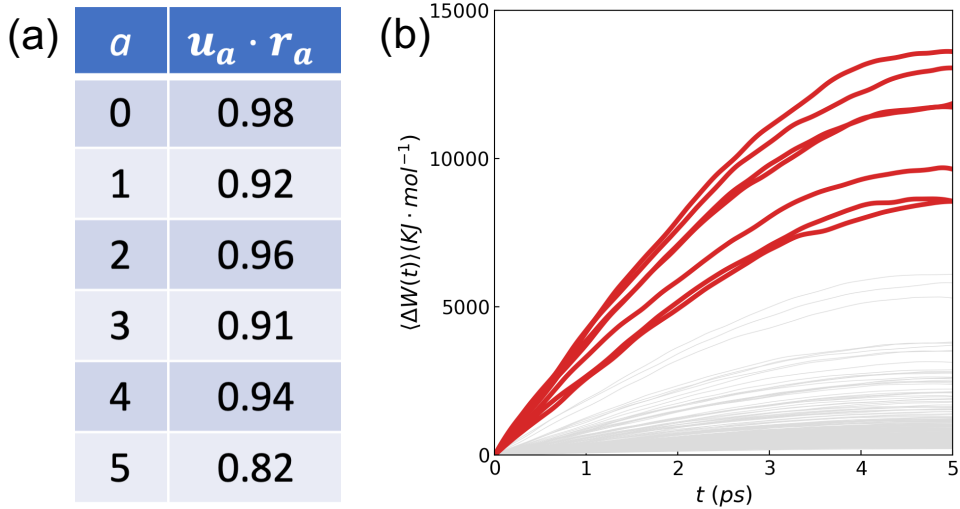

**Fig. 2: Singular vectors of energy relaxation of HIV-PR**. **(a)** Inner product between leading SCs ($u_a$) of energy relaxation and tRCs ($r_a$) of flap opening of HIV-PR in implicit solvent. **(b)** Potential energy flows of SCs of DRV-bound HIV-PR averaged over $E_0(u_0)$ (details in Methods). Red lines represent $u_0$ to $u_6$; gray lines represent the other SCs (i.e. non-RCs).

Figure 2a shows the inner product between the six leading SCs for the energy relaxation of HIV-PR in implicit solvent and the six tRCs of its flap opening we identified in ref. [34]. Each SC ($u_a$) or tRC ($r_a$) is a linear combination of the backbone dihedrals ($\chi_i$) of HIV-PR: $u_a = \sum_{a=1}^{N} U_{ia}\chi_i$. All inner products are close to 1—confirming our hypothesis that the leading SCs of energy relaxation are the same as the tRCs for conformational transitions.

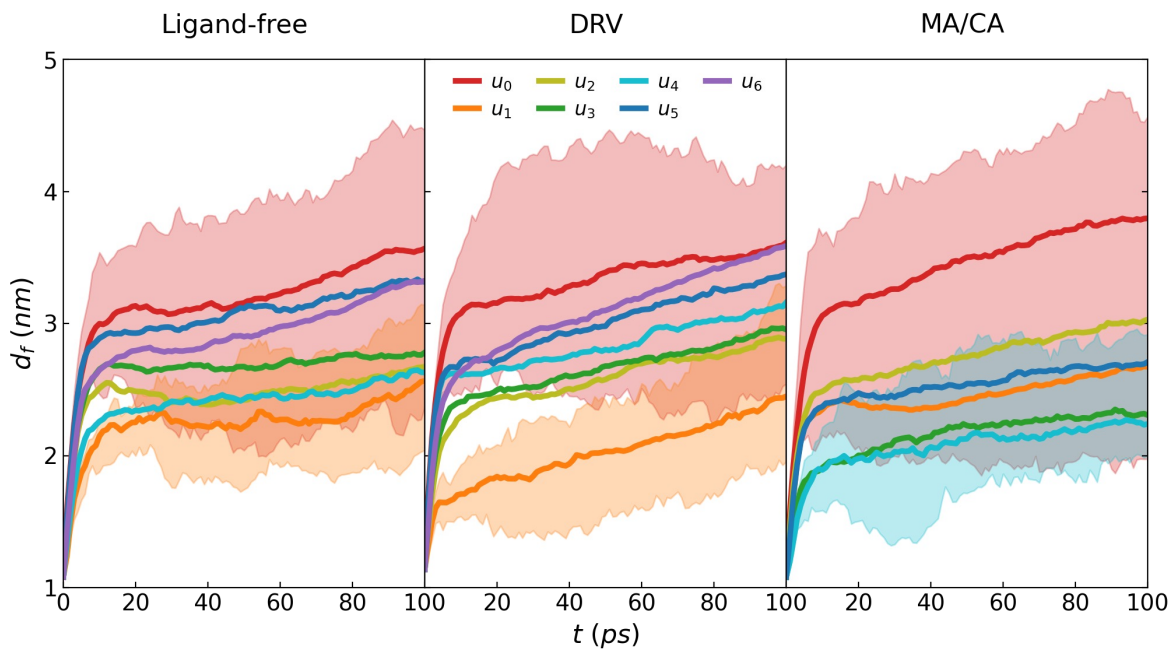

**Fig. 3: Efficiency of flap opening along RC-uncovered trajectories in ligand-free, DRV-bound and MA/CA-bound HIV-PR.** Solid lines are results of averaging over 30 trajectories. Each shaded region shows the range of $d_f$ covered by trajectories generated from the corresponding SC. The upper bound is the maximum and the lower bound is the minimum $d_f$ value reached at each time point among the 30 trajectories used in calculating the average.

This result revealed a fundamental connection between activation and energy relaxation, which, in our understanding, stems from the need for optimal efficiency in protein function. Both processes rely on systematic protein dynamics driven by systematic PEFs through individual coordinates, which are absent during equilibrium fluctuations. The importance of a coordinate is determined by the magnitude of its PEF, which is an intrinsic physical property determined by the protein structure and fine-tuned through evolution. tRCs, with the highest PEFs, serve as optimal energy flow channels.

During activation, the protein delivers energy to the active site to perform its function. For optimal delivery, energy systematically flows from low-capacity channels (non-RCs) to high-capacity channels (tRCs). During energy relaxation, the protein dissipates excess energy from the active site to the rest of the protein and eventually to the solvent. For efficient dissipation, energy first flows into the tRCs for rapid removal and then disperses into the non-RCs. Thus, energy flows through the same channels in both activation and relaxation but in opposite directions. This dual role of tRCs allows us to compute them from energy relaxation as well as from activated processes.

**(a) DRV unbinding (RC-uncovered trajectory)**

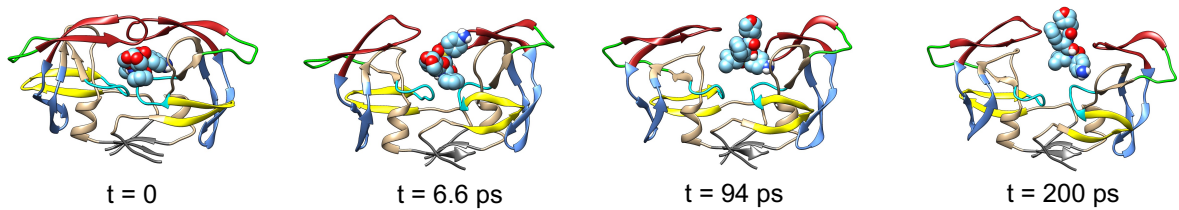

t = 0　　　　　　　t = 6.6 ps　　　　　　t = 94 ps　　　　　　t = 200 ps

**(b) DRV unbinding (NRT)**

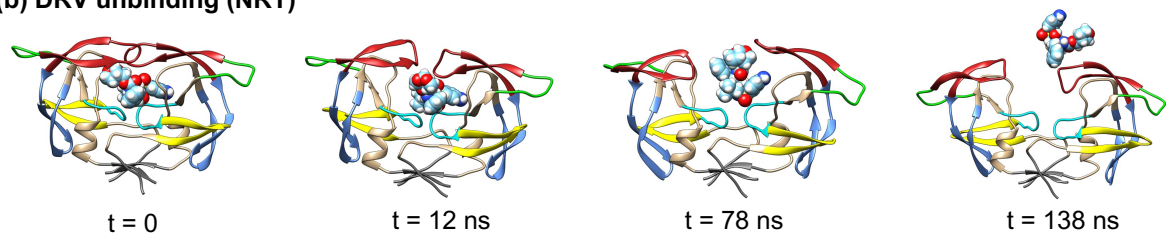

t = 0　　　　　　　t = 12 ns　　　　　　t = 78 ns　　　　　　t = 138 ns

**(c) MA/CA unbinding (RC-uncovered trajectory)**

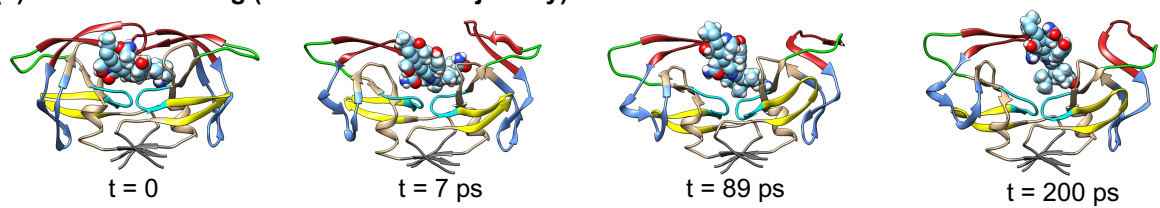

t = 0　　　　　　　t = 7 ps　　　　　　　t = 89 ps　　　　　　t = 200 ps

**(d) MA/CA unbinding (NRT)**

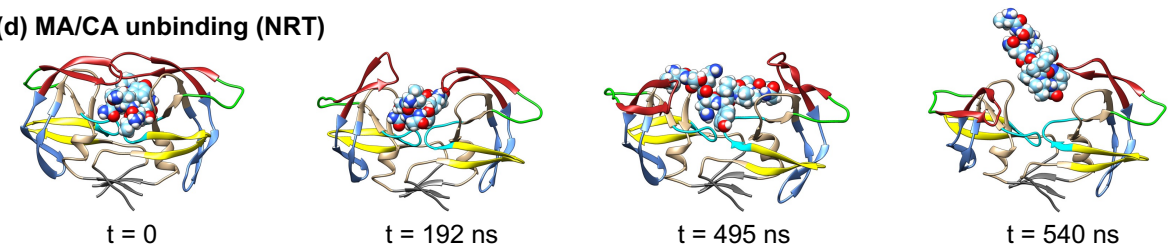

t = 0　　　　　　　t = 192 ns　　　　　　t = 495 ns　　　　　　t = 540 ns

**Fig. 4**: **Snapshots from RC-uncovered trajectories and NRTs of flap opening and ligand unbinding in HIV-PR bound to DRV (a,b) and (MA/CA peptide (c, d)**. The full trajectories are shown in Supplementary videos S1 to S4, respectively.

**tRCs provide effective enhanced sampling of conformational changes.** A common assumption is that tRCs are the optimal CVs for enhanced sampling of protein conformational changes [8]. If we can compute tRCs from energy relaxation, it will be highly valuable for applications. To explore this possibility, we simulated energy relaxation in two systems, HIV-PR bound with DRV and MA/CA peptide in explicit solvent. Figure 2b shows the potential energy flows of SCs of energy relaxation of DRV-bound HIV-PR. The first seven SCs ($u_0$ to $u_6$; red lines in Fig. 2b) show much higher potential energy flows than the rest (gray lines in Fig. 2b), suggesting they are the tRCs.

The same holds true for the SCs of MA/CA-bound HIV-PR. To distinguish, we will refer to them as tRCs from now on.

By applying bias potentials on the tRCs (details in Methods section and ref. [34]), we generated RC-uncovered trajectories for flap opening and ligand unbinding of HIV-PR in ligand-free, DRV-bound, and MA/CA-bound states. The time evolution of the flap distance $d_f$ (Fig. 1a) along these trajectories is shown in Fig. 3. For all tRCs and across all systems, $d_f$ quickly increases to the range of 2.5 to 3 nm within 20 ps, corresponding to wide-open conformations in the literature [46,47]. Moreover, the ligand dislodges from the active site and is well on its way to exit the flaps within 200 ps (Fig. 4a,c; Supplementary video S1, S2).

To evaluate the efficiency of enhanced sampling by the tRCs, we compare the time scales of flap opening and ligand unbinding in RC-uncovered trajectories with those observed in other simulations and experiments [47-49]. Sadiq and Fabritiis conducted 461 MD trajectories, each 50 ns long, totaling 23 μs simulation time [47]. They observed four wide-open events in ligand-free HIV-PR, averaging 5.8 μs per event, in line with the 100 μs observed by NMR [46]. In comparison, RC-uncovered trajectories show wide-open flaps within 10 ps, at least a $2.9 \times 10^5$-fold acceleration compared to MD for ligand-free HIV-PR. For MA/CA-bound HIV-PR, Sadiq et al conducted 100 MD simulations totaling 10 μs and observed only thermal fluctuations [49]. Thus, flap opening in this system requires much longer than 10 μs. In stark contrast, RC-uncovered trajectories demonstrate wide-open flaps within 20 ps, marking an acceleration far exceeding $10^6$-fold. Additionally, while experimentally measured half-life of DRV unbinding is $8.9 \times 10^5$ s [50], RC-uncovered trajectories achieved this in 200 ps (Fig. 4a). Together, these results demonstrate the efficiency of RC-uncovered trajectories.

**Reversible transitions along individual tRCs**. Beyond sampling the transition paths, an important goal of enhanced sampling is to compute thermodynamic variables, especially free energy surfaces. This requires that the system undergo reversible transitions. The current system has multiple tRCs, with different tRCs opening the flaps in distinct ways, as shown in Fig. 6 in ref. [34]. Therefore, it is crucial that the flaps can open and close reversibly along a single tRC so that, with proper application of bias potentials on multiple tRCs, a multi-dimensional free energy surface can be computed. Figure 5 shows time evolution of both $u_0$ and the flap distance along an RC-uncovered trajectory obtained from applying bias on $u_0$ of ligand-free HIV-PR, featuring multiple rounds of flap opening and closing. The high similarity in time dependence of both $\Delta u_0(t)$ and $d_f(t)$ across different transition cycles demonstrates the robustness of transition reversibility.

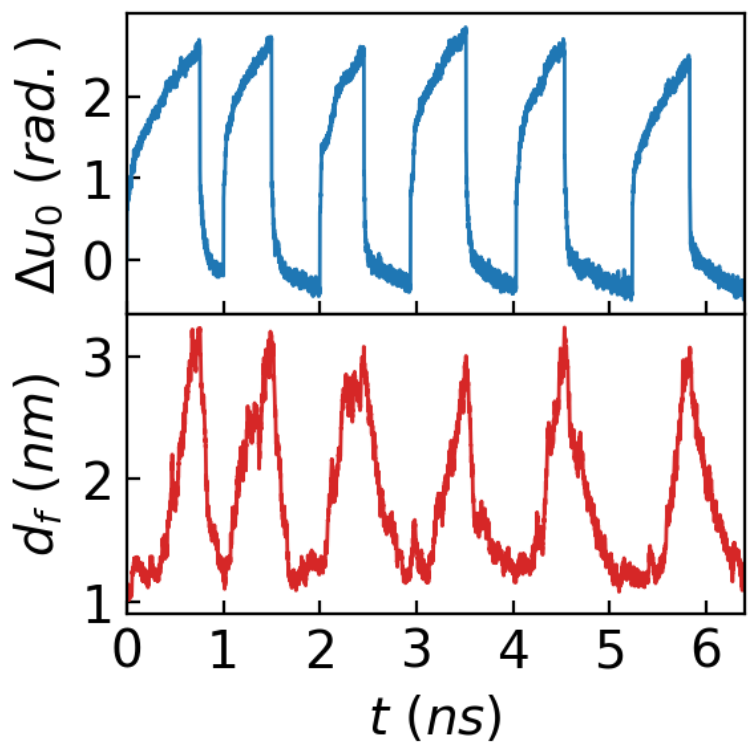

**Fig. 5: RC-uncovered trajectory of HIV-PR with reversible transitions.** Time evolution of tRC $u_0$ (blue) and flap distance $d_f$ (red) along an RC-uncovered trajectory of $u_0$ with multiple transitions between close and open states.

**RC-uncovered trajectories follow natural transition pathways.** A major advantage of RC-uncovered trajectories is that they follow natural transition pathways, the same pathways traced by NRTs. This is crucial for understanding protein functions, the main goal of MD simulations. A rigorous way to verify whether a trajectory follows natural pathways is through the committor. Along an NRT, the committor covers the full range of values: $p_B \in [0, 1]$, with $p_B \in (0.1, 0.9)$ marking barrier crossing. By contrast, a trajectory deviating from natural pathways will show an abrupt jump in committor values from 0 to 1 or do not manifest well-defined committor values, as it bypasses the actual activation barrier and TS. Instead, it explores a non-physical region of the conformational space, where committor values are not well-defined. Therefore, a progression through intermediate committor values is a clear signature that a trajectory follows natural transition pathways.

In Fig. 7a of ref. [34], RC-uncovered trajectories of all 6 tRCs span the full range of committor values, confirming that they follow natural transition pathways. Therefore, an RC-uncovered trajectory that follows natural pathways can validate the corresponding RC as a tRC. For HIV-PR in explicit solvent, calculating committor is prohibitively expensive. As a computationally efficient but rigorous alternative, we use the shooting move of TPS [10] to show that RC-uncovered trajectories of the current systems pass through the TS region. The TS is the bottleneck along a natural transition pathway and represents the most critical intermediate committor value: $p_B \simeq 0.5$. It is difficult to envision a scenario where a trajectory reaches the TS but still diverges from the natural pathway.

In the shooting move [10], we select a conformation $\boldsymbol{R}_0$ from an RC-uncovered trajectory and draw momenta $\boldsymbol{p}_0$ from Boltzmann distribution. We then launch a pair of MD trajectories from $\boldsymbol{R}_0$ with initial momenta $\boldsymbol{p}_0$ and $-\boldsymbol{p}_0$, respectively. If these trajectories reach opposite basins, we leverage the time reversibility of classical mechanics to create a reactive trajectory by reversing the momenta along the trajectory that ends in the reactant basin and merging the two trajectories at $\boldsymbol{R}_0$ [10]. Since no bias is used, trajectories generated by the shooting move are NRTs, as demonstrated by Chandler and colleagues [10,11].

The likelihood of successfully generating NRTs from a conformation $\boldsymbol{R}_0$ is determined by $p(RT|\boldsymbol{R}_0)$, the probability that a dynamic trajectory passing through $\boldsymbol{R}_0$ is an NRT. This probability is related to the committor by $p(RT|\boldsymbol{R}_0) = 2p_B(\boldsymbol{R}_0)\big(1 - p_B(\boldsymbol{R}_0)\big)$ [51,52], which reaches a maximum of 0.5 at $p_B(\boldsymbol{R}_0) = 0.5$, and drops to 0 at $p_B(\boldsymbol{R}_0) = 0$ or 1. Given this, the likelihood of generating NRTs via shooting is significant only when $\boldsymbol{R}_0$ is close to the TS—where the free energy difference between $\boldsymbol{R}_0$ and the TS is within the thermal energy $k_B T$. Successfully generating NRTs through the shooting move provides objective validation that RC-uncovered trajectories pass through the TS, thereby confirming that they follow natural pathways.

We obtained NRTs for flap opening and complete ligand dissociation in DRV-bound and MA/CA-bound HIV-PRs (Fig. 4b, d; Supplementary videos S3, S4) by shooting from the respective RC-uncovered trajectories. To our knowledge, this is the first successful attainment of NRTs for ligand dissociation from HIV-PR. For computational efficiency, only five pairs of shooting trajectories were attempted per conformation. Consequently, conformations that successfully generate NRTs are likely close to the TS. These results demonstrate that the RC-uncovered trajectories pass through TS and follow the natural pathways, validating the tRCs computed from energy relaxation simulations.

**Comparison with enhanced sampling by empirically derived CVs**. tRCs are widely recognized as the optimal CVs for enhanced sampling of protein conformational changes. To test this, we simulate MA/CA dissociation from HIV-PR by applying bias potentials on commonly used empirical CVs and compare the results with NRT and RC-uncovered trajectories. The NRT provides the correct mechanism of this process because it is an unbiased MD trajectory that covers the entire transition period.

The empirical CV we use is the distance $s_l$ between the center of mass (CoM) of $C_\alpha$ atoms of MA/CA and the CoM of $C_\alpha$ atoms of active site residues (residues 24, 26, 27). We chose $s_l$ because it was employed in ref. [17], an extensive metadynamics-based bias-exchange simulation of the dissociation of a six-residue peptide from HIV-PR that has an accumulated simulation time of 1.6 μs.

We first present the results from steered MD simulation using the umbrella pulling method implemented in GROMACS, similar to how we applied bias to tRCs. The pulling strength on $s_l$ was adjusted to achieve dissociation in 1.4 ns. As shown in Fig. 6, the resulting trajectory is fundamentally different from the NRT and RC-uncovered trajectory (Fig. 4c,d).

In the steered MD trajectory (Fig. 6a), $s_l$ and the flap distance $d_f$ increase together, indicating that the ligand is actively pushing the flaps open, causing elastic distortion of the flaps (Fig. 6d). Moreover, $d_f$ only reaches 1.8 nm, just enough for the ligand to slip through. We consider it an elastic distortion rather than a true flap-opening because it involves only the bending of the flaps (Fig. 1). This behavior contradicts the established mechanism of HIV-PR function, which requires full flap-opening.

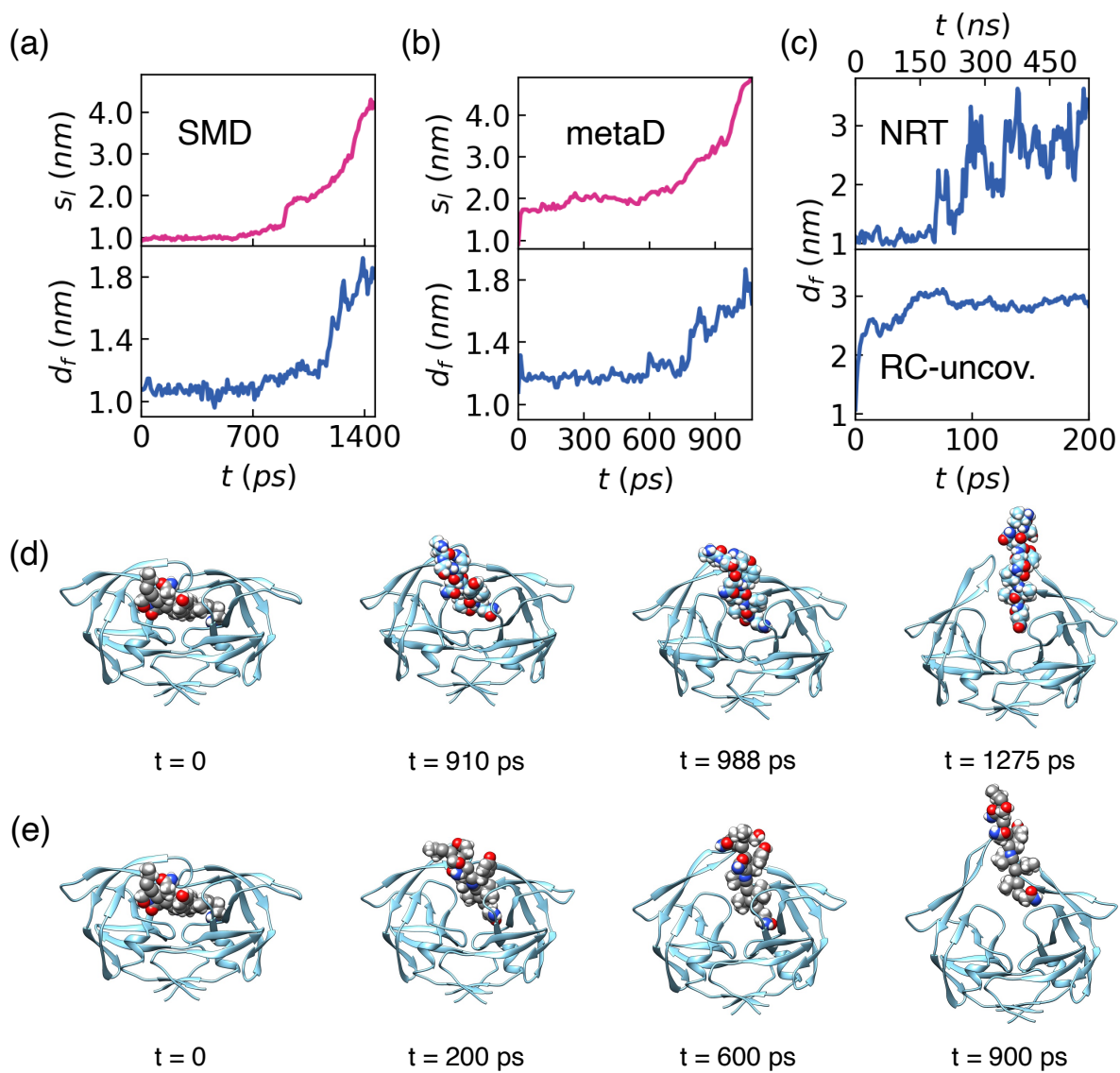

**Fig. 6: Non-physical features of enhanced sampling with empirical CVs**. (**a**) Time evolution of $s_l$ and $d_f$ along a steered MD trajectory of MA/CA dissociaiton from HIV-PR in explicit solvent with $s_l$ as the CV. (**b**) Time evolution of $s_l$ and $d_f$ along a metadynamics trajectory. (**c**) Time evolution of $d_f$ along an NRT and an RC-uncovered trajectory. (**d**) Snapshots from the steered MD trajectory in (a). (**e**) Snapshots from the metadynamics trajectory in (b)

By contrast, in both NRTs and RC-uncovered trajectories (Fig. 4), the flaps open to 2 nm as the ligand begins to dislodge (192 ns for the NRT, 7 ps for the RC-uncovered trajectory). Afterwards, the influx of water molecules into the active site drives ligand motion and further flap opening.

Full ligand dissociation occurs when the flaps open to 3 nm. Furthermore, flap opening in NRT and RC-uncovered trajectory (Fig. 4) is driven by global, collective protein structural changes, causing the flaps to swivel open rather than just bend. The main difference between the NRT and RC-uncovered trajectory is the longer time scale and larger fluctuations in $d_f$ along the NRT.

These results demonstrate that the NRT and RC-uncovered trajectory follow the same transition pathway, which aligns with the established mechanism of ligand dissociation from HIV-PR and contrasts sharply with the non-physical behavior seen in the steered MD trajectory, a typical issue when applying bias on non-RC CVs.

For fair comparison, we need to ensure that the non-physical features shown in Fig. 6d are due to the empirical CV, not the specifics of how the bias is applied. For this purpose, we simulated the same process using the same CV with the well-tempered metadynamics implemented in PLUMED2 [53]. Unlike steered MD, metadynamics applies bias potentials in a highly adaptive and flexible way, minimizing the risk of artifacts from the non-adaptive bias protocol used in steered MD. Figure S4 shows the time evolution of $d_f$ and $s_l$ under varying strength of Gaussian bias. To achieve flap-opening within 2 ns, a Gaussian height at 10,000-fold the recommended value (0.2 kJ mol$^{-1}$) is needed. The resulting trajectory displays the same features as the steered MD simulation, evident in the time evolutions of $s_l$ and $d_f$ (Fig. 6b) and the snapshots (Fig. 6e). Notably, ref. [17] reported peptide dissociation with minimal flap opening, in line with our observations here. These results confirm that the non-physical characteristics of the MA/CA dissociation trajectories are the consequence of using $s_l$ as the CV.

**Efficiency of enhanced sampling by tRCs**. In the results above, the time scales for DRV dissociation in RC-uncovered trajectories, NRTs, and experiments differ by orders of magnitude. This is because they emphasize different phases of the same process. The experimental half-life corresponds to the average time for observing a DRV dissociation event in an extremely long MD trajectory containing many rounds of DRV binding and unbinding. In this context, NRTs represent a $10^{12}$-fold acceleration over MD. This is enabled by the shooting move's focus on the actual barrier crossing phase while bypassing the extended waiting time in the reactant basin, which is the determinant of the half-life of an activated process [10,11]. To bypass waiting, shooting move must start from a TS conformation—the least populated conformational state that requires a long waiting period to reach—which is provided by RC-uncovered trajectories. In contrast, enhanced sampling trajectories obtained with empirically derived CVs do not sample TS conformations. Finally, RC-uncovered trajectories are ~$10^3$-fold shorter than NRTs, aggregating into a $10^{15}$-fold acceleration over MD. This is achieved by reducing diffusive motions—the most time-consuming aspect of NRTs [54].

The efficiency of RC-uncovered trajectories is enabled by an intriguing physical mechanism. The critical step of an activated process is energy activation, where rare fluctuations channel energy into the tRCs, enabling the system to cross the activation barrier [30,31]. The efficiency of enhanced sampling hinges upon effective transfer of energy into the tRCs to expedite activation. tRCs are optimal CVs because, in this case, the bias potentials directly inject energy into tRCs, thereby maximizing the efficiency of energy activation.

**Predictive enhanced sampling by tRCs: PDZ domain allostery.** Energy relaxation requires only a single protein structure, yet the resulting tRCs enable enhanced sampling of the protein's inherent conformational transitions, underscoring the predictive capability of our approach. In our HIV-PR simulations, the only input was the flap-closed conformation, but we successfully obtained NRTs of the flap opening process and the ensemble of flap-open conformations. To further validate this predictive capability, we use the PDZ domain as a blind test, because it has only one known structure, and its allostery has been a challenging puzzle for over two decades.

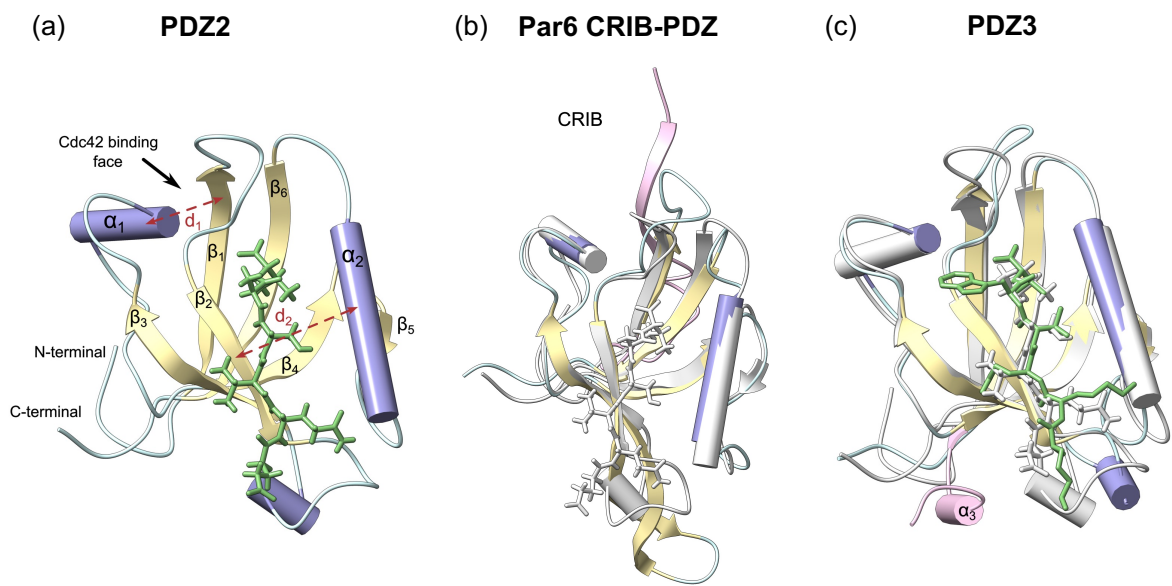

**Fig. 7: Structures of PDZ domains and allosteric sites**. **(a)** Structure of PDZ2 domain (PDB: 3LNY) bound to 6-residue (EQVSAV) peptide. Different structural elements labeled. **(b)** Structure of Par6 CRIB-PDZ domain (PDB: 1NF3) aligned to PDZ2 (White). The CRIB domain is colored Pink. **(c)** Structure of PDZ3 domain (PDB: 1TP5) bound to a 6-residue (KKETWV) peptide aligned to PDZ2 (White). The $\alpha_3$-helix is colored Pink.

PDZ domains are a large family of protein-ligand interaction modules [55]. They share the same canonical fold (Fig. 7) and function as organization centers in multi-protein signaling complexes. Two examples of PDZ allostery have been extensively studied. The first is a ~13-fold increase in ligand binding affinity when Cdc42 binds to the Par6 CRIB-PDZ (Fig. 7b) [56] at the interface formed by $\alpha_1$-helix, and $\beta_1$-sheet. The second example involves an extra $\alpha_3$-helix appended to the C-terminal of PDZ3 (pink helix in Fig. 7c). Two studies found a consistent conclusion [57]: the presence of $\alpha_3$ increases PDZ3 ligand affinity by 21- to 120-fold, despite the lack of direct interactions between $\alpha_3$ and the ligand binding groove between $\alpha_2$ and $\beta_2$ (Fig. 7a) [58].

The difficulty in understanding PDZ allostery is that crystal structures across different PDZ domains in both ligand-bound and ligand-free states are virtually identical (Fig. 7) [55], challenging the conventional view that allosteric effectors modify ligand affinity by exploiting the structural differences between apo and holo states. It is difficult to understand how the perturbations introduced by effectors—Cdc42 or $\alpha_3$—at distal sites are communicated to the binding groove and alter ligand affinity. Despite intensive efforts for over two decades, the mechanism of PDZ allostery remains elusive. This ongoing challenge is encapsulated by the title of a recent review: Allostery Frustrates the Experimentalist [44].

The process underlying PDZ domain allostery is its ligand binding; it has been speculated that conformational changes during this process are responsible for allostery [59,60]. A solid understanding of ligand binding could resolve the PDZ puzzle unequivocally. However, the time scale for PDZ ligand unbinding—estimated to be 10 to 100 ms—is too slow for MD [59]. Consistent with this estimation, in an extensive MD simulation that totaled 0.5 ms, no ligand unbinding or discernable conformational change in PDZ2 was observed [59]. Therefore, ligand unbinding from PDZ2 provides an ideal blind test for our method's predictive capability.

**Mechanism of PDZ allostery from predictive enhanced sampling.** Figure S5a shows the potential energy flows along SCs from energy relaxation of PDZ2 bound to an eight-residue (ENEQVSAV) RA-GEF2 peptide [42]. There is a clear gap between the first six SCs ($u_0$ to $u_5$) and the rest, suggesting they are the tRCs. To distinguish, we refer to them as the tRCs from now on.

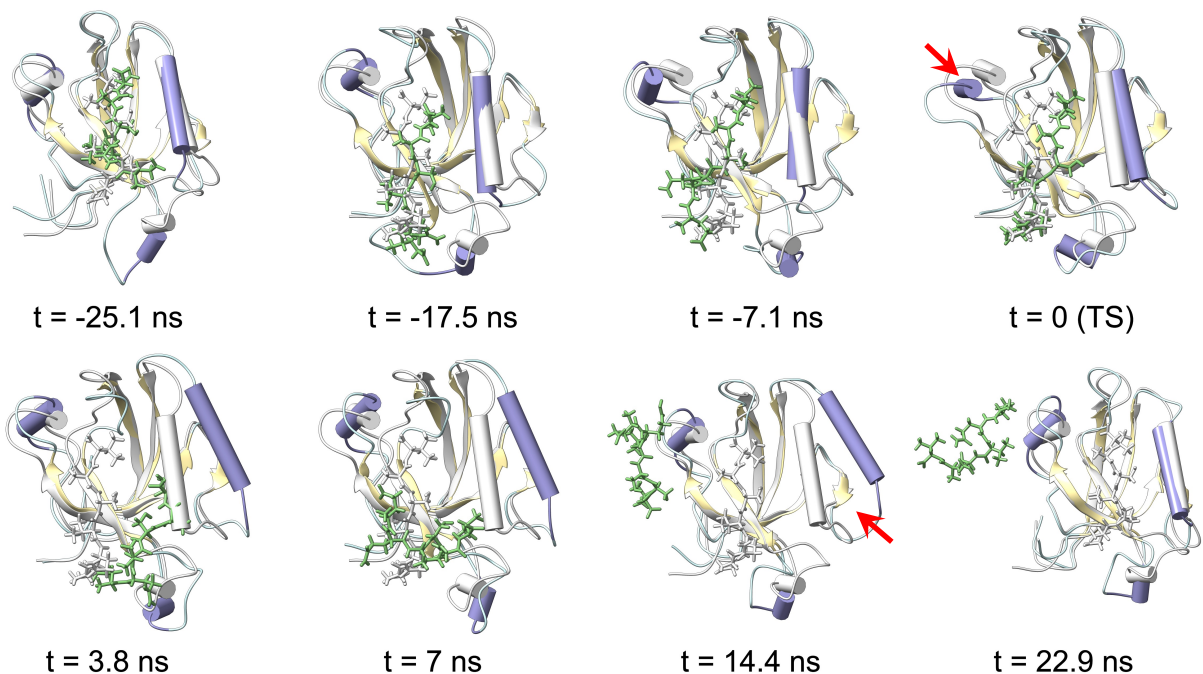

**Fig. 8: Snapshots from a natural trajectory for PDZ2 ligand unbinding.** The crystal structure of PDZ2 is shown in light grey as reference to illustrate the transient nature of the conformational changes during ligand binding. Red arrows point to locations of large-scale conformational changes.

RC-uncovered trajectories show significant PDZ2 conformational changes and ligand dislocation from the binding groove. Figure S5b presents end structures along RC-uncovered trajectories of the six RCs. They show a consistent pattern: the ligand-binding site opens between $\alpha_2$ and $\beta_2$, while the Cdc42 binding face spanning over the $\alpha_1$-$\beta_1$ cleft significantly expands due to the large shift of $\alpha_1$ and the $\alpha_1$-$\beta_4$ loop (Fig. 8, 9). These results strongly suggest that PDZ allostery is due to effectors interfering with conformational changes critical for ligand binding. To validate this hypothesis and pin down the allosteric mechanism, we generate NRTs (Fig. 8) for ligand unbinding by applying the shooting move on TS conformations from RC-uncovered trajectories [10,11]. Figure 9b shows example TS conformations, which are identified by their ability to generate NRTs. They all show the opening of the $\alpha_1$-$\beta_1$ cleft and the binding groove, highlighting the critical importance of these changes to ligand unbinding. Across the TS ensemble, opening of the binding groove is virtually identical, whereas opening of the $\alpha_1$-$\beta_1$ cleft span a small range marked by the TS conformations in cyan and yellow in Fig. 9b.

The NRT for ligand unbinding in Fig. 8 (video S6) shows a transient large-scale conformational change in PDZ2 that lasts 20 to 30 ns. This conformational change shares the same features shown in RC-uncovered trajectories: opening in both the $\alpha_1$-$\beta_1$ cleft and the binding groove. These two conformational changes are concerted. Figure 9a shows the time evolution of three CVs for characterizing PDZ2 conformational changes and ligand dissociation along the NRT in Fig. 8. The distance $d_1$ between the CoMs of $\alpha_1$ and upper $\beta_1$ marks the opening of the $\alpha_1$-$\beta_1$ cleft, $d_2$ between the CoMs of $\alpha_2$ and $\beta_2$ characterizes the opening of the binding groove, and $d_3$ between CoMs of the ligand and the binding groove delineates ligand dissociation. During ligand unbinding, the $\alpha_1$-$\beta_1$ cleft opens first, marked by the increase of $d_1$ from 8 Å in the crystal structure to 12 Å in the TS. After reaching the TS, $d_1$ starts to decrease while $d_2$ starts to increase from its value of

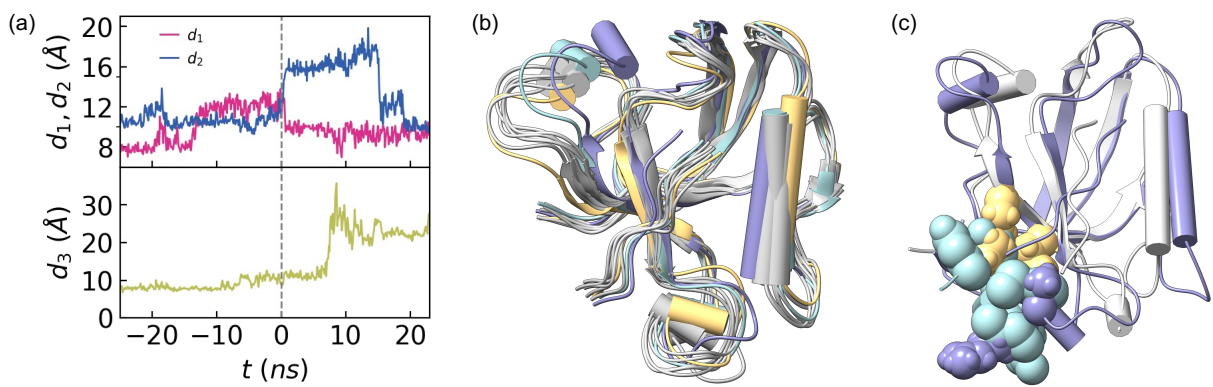

**Figure 9: Conformational dynamics of PDZ2 during ligand unbinding. (a)** Time evolution of the three parameters $(d_1, d_2, d_3)$ we used to characterize PDZ2 conformational change along the trajectory shown in Fig. 6. TS occurs at time 0 for easy comparison. **(b)** TS ensemble (Gray) harvested from RC-uncovered trajectories. The crystal structure is shown in Purple as comparison. TS conformations in Yellow and Cyan show the largest and smallest conformational changes among all the TS we harvested. The ligands are shown in cartoon representation. **(c)** Steric clashes in the TS conformation of the trajectory in Fig. 6. Crystal structure of PDZ3 is shown in White and TS is shown in Purple. Residues involved in the steric clashes are shown in spheres. Residues in $\alpha_3$ of PDZ3 are in Cyan. Residues Tyr36 and Lys38 of PDZ2 are in Yellow; Val30 and Arg31 are in Purple.

10 Å in the crystal structure until it reaches 16 Å. The period of $d_2$ increase coincides with the dislodging of the ligand, which begins to dissociate rapidly from the binding groove once $d_2$ reaches 16 Å. After the ligand fully dissociates, both the $\alpha_1$-$\beta_1$ cleft and the binding groove return to their crystal structure conformation in 10 ns. As shown in Fig. 8, the opening of the $\alpha_1$-$\beta_1$ cleft and the binding groove occur only during the barrier crossing process, leaving the PDZ2 conformation at the beginning and the end of the NRT effectively the same as the crystal structure. The transient duration of this critical conformation change explains why it has never been observed in experiments, demonstrating the unparalleled value of NRTs in providing mechanistic insights into protein function.

The $\alpha_1$-$\beta_1$ cleft is the major component of the Cdc42 binding interface (Fig. 7b), thus Cdc42 binding will hinder its opening. As shown in Fig. 9c, opening of the $\alpha_1$-$\beta_1$ cleft moves the $\alpha_1$-$\beta_3$-$\beta_2$ block toward $\alpha_3$, while opening the binding groove moves the $\beta_2$-$\beta_3$ loop toward $\alpha_3$. The combined effects are the severe steric clashes in the TS (Fig. 9c): between $\alpha_3$ and two residues (Tyr36, Lys38) in $\beta_3$, and between $\alpha_3$ and two residues (Val30, Arg31) in the $\beta_2$-$\beta_3$ loop. These results suggest that both Cdc42 and $\alpha_3$ modify ligand affinity to PDZ by interfering with the transient conformational changes during ligand unbinding, potentially slowing down the dissociation process. This interaction provides a straightforward and intuitive mechanism for PDZ allostery. Overall, our simulation results demonstrate the predictive capability of our method by enabling the detection of large-scale conformational changes critical for PDZ allostery that eluded intensive studies for two decades [44].

**Discussions**

In this work, we discovered that tRCs control both activation and energy relaxation, revealing a surprising reciprocity between these processes despite their significant differences in energy, timescale, and motion. Energy dissipated in relaxation is on the order of thermal energy, whereas activation requires energy far beyond thermal levels. Additionally, while energy relaxation occurs within a few picoseconds, ligand unbinding in HIV-PR can take over 200 hours. Energy relaxation is also characterized by small-amplitude vibrational motions, in contrast to the large-scale conformational changes typical of activation. These differences highlight the complexity of the underlying mechanisms. Yet, both processes are governed by the same set of essential coordinates—the tRCs—demonstrating a fundamental unity between these seemingly disparate phenomena. If this reciprocity is confirmed across other proteins, it could lead to a generalization of the fluctuation-dissipation theorem in proteins, extending its applicability to macroscopic energy scales.

Our finding has significant applications in enhanced sampling, the primary method for bridging the time-scale gap between MD simulations and functionally important protein processes. The main challenge in enhanced sampling is identifying CVs that can effectively accelerate protein conformational changes. While tRCs are widely regarded as the optimal CVs, their identification previously depended on NRTs, which themselves required effective enhanced sampling, creating a paradox. Our discovery allows for the computation of tRCs from energy relaxation at minimal computational cost, enabling predictive and efficient enhanced sampling of protein conformational changes. This significantly broadens the range of protein functional processes accessible to MD simulations.

Although the concept of tRCs applies only to single-step reactions, our method can be adapted for multi-step reactions by following the standard chemical kinetics approach of dividing them into a sequence of single-step reactions. Starting from a known conformation, we can obtain tRCs for transitions into neighboring conformations, enabling simulations of these transitions and prediction of the resulting conformations (Fig. S6). This procedure can then be repeated in a stepwise manner to reach conformations further from the starting point.

The GWF is a general, flexible method that can identify highly curved tRCs through piecewise linearization, as detailed in ref. [43]. In practice, for all processes we have studied, the tRCs have been linear within numerical error [34,43]. It remains to be seen whether this linearity is a general feature of tRCs in proteins and if there is a fundamental physical reason for it.

**Methods**

All simulations are constant NVE and use the CHARMM36m force field and TIP3P water model [61,62]. For HIV-PR, the simulation system consists of 71,589 atoms from water and 3280 atoms from protein and ligand. For PDZ2, the simulation system consists of 22,101 atoms from water and 1482 atoms from protein and ligand. Simulations of energy relaxation and RC-uncovered trajectories are performed with GROMACS 2019.2 using CPUs, whereas natural trajectories are generated using GROMACS 2022 on GPUs [63]. All the bonds involving hydrogen atoms are constrained using the LINCS algorithm [64]. Time step used in all simulations is 1 fs.

To simulated energy relaxation of HIV-PR in explicit solvent, extra kinetic energy is deposited into the ligand to increase the temperature of each ligand coordinate by 400K, the same as we did for myoglobin [42]. For energy relaxation of ligand-free HIV-PR in implicit solvent, extra kinetic energy was injected into atoms of active site residues (residues 25, 28, 29) instead. This is to mimic the process of dissipating the excess energy at the active site after ligand binding or enzymatic reaction. To test if the SCs depend on the specific amount of excess energy deposited into the ligand, we also tried increasing the ligand temperature by 150K and 800K. The results are the same. For the GWF analysis, the 8,000 ER trajectories are divided into 6 clusters ($E_0(q_i)$ to $E_5(q_i)$) and the GWF was computed using cluster $E_0(q_i)$, which contains 3,300 trajectories (details in ref. [42] and Supplementary Information). The detailed procedure for simulating ER and conducting the GWF analysis is the same as used for myoglobin in ref. [42] and will not be repeated here.

To generate RC-uncovered trajectories, we apply bias potentials to tRCs $u_a$ ($a = 0,1,…,6$). Because of the periodicity of angles, we replace $u_a$ by $Q_a = \sum_{i=1}^{N_a} U_{ia} \cos(\chi_i - \chi_i^*)$ [65], where $U_{ia}$ is an element of the left singular matrix of the GWF, $N_a$ is the number of coordinates included in the definition of $Q_a$, $\chi_i$ denotes a dihedral, $\chi_i^* = \chi_i^0 + c \cdot U_{ia}$ is the target value of $\chi_i$, $\chi_i^0$ is the value of $\chi_i$ in the starting structure, and $c$ is the constant that we use to decide how many units of $u_a$ we want to move the system. The use of cosine function is to remove the discontinuity caused by the periodicity of angular coordinates. Because the GWF inevitably contains noise, we do not include all the backbone dihedrals in $Q_a$. Instead, we only include $\chi_i$ if $|U_{ia}| > \varepsilon$ to remove dihedrals that are included in $u_a$ due to the noise in the GWF. In current calculations, we used $\varepsilon = 0.03$, resulting in $N_a \in (240, 290)$ for different RCs of HIV-PR.

To apply the bias force gradually, we divide the interval between the minimum value $-c \sum_i U_{ia}$ and the target value $c \sum_i U_{ia}$ of $Q_a$ into 100 bins, with the n-th bin spanning the interval $[Q_{a,n}, Q_{a,n+1})$. At any instant $t$, when the system configuration is at $Q_a(t) \in [Q_{a,n}, Q_{a,n+1})$, it feels an bias potential of the form $V(Q_a) = \frac{1}{2} k (Q_a(t) - Q_{a,n+1})^2$. In this way, the system always feels a gentle force pulling it towards the target value, with the center of the bias potential shifting adaptively as the system moves from one bin to another. Even though the relationship between $Q_a$ and $u_a$, Fig. 5 shows that applying a bias potential to $Q_a$ moves $u_a$ continuously and efficiently, suggesting that the bias force on $Q_a$ is effectively translated into a force acting on $u_a$. Because $Q_a$ is a unitless variable, the spring constant $k$ has a unit of kJ mol$^{-1}$.

To achieve multiple rounds of reversible transitions, we will reverse the direction of the bias potential once the system reaches the terminal bins centered at $\Delta u_0 = 0$ and 2.6 radians and has stabilized there for 50 ps. To balance the duration for flap opening and closing transitions and minimize the overall duration of each cycle, we empirically selected $k = 2000$ kJ mol$^{-1}$ for flap opening and $k = 500$ kJ mol$^{-1}$ for flap closing, respectively.

To generate NRTs using the shooting move [10], we first identify a set of candidate TS conformations from an RC-uncovered trajectory based on visual inspection and intuition. From each candidate conformation, we launch 5 pairs of MD trajectories with opposite initial momenta. A natural reactive trajectory is successfully generated when a pair of MD trajectories launched from a candidate conformation reached opposite basins. This procedure also allows us to identify TS conformations by their success in generating natural trajectories.

For metadynamics simulation of MA/CA dissociation, we used the widely adopted well-tempered metadynamics implemented in PLUMED2 [66]. The parameters used in the simulations are: Gaussian height 0.2 kJ/mol, Gaussian width 0.25 nm, and deposition of a Gaussian every 0.5 ps, all based on PLUMED2 recommendations and ref. [67]. Additional Gaussian heights of 1, 100, 1000, and 2000 kJ/mol are also tested (Fig. S4) to achieve MA/CA dissociation.

**Data Availability**

All data generated or analyzed during this study are included in this published article and its Supplementary Information file.

**Code Availability**

All the custom codes used in this study are deposited in Code Ocean.

**Author Contributions**

A.M. designed the research. H.L. and A.M. conducted research and analyzed data. A.M. wrote the manuscript.

**Competing Interests**

The authors declare no competing interests.

**References**

1. Frauenfelder, H., Sligar, S. G. & Wolynes, P. G. The energy landscapes and motions of proteins. *Science* **254**, 1598-1603. (1991).
2. Henzler-Wildman, K. & Kern, D. Dynamic personalities of proteins. *Nature* **450**, 964-972, doi:10.1038/nature06522 (2007).
3. Xie, T., Saleh, T., Rossi, P. & Kalodimos, C. G. Conformational states dynamically populated by a kinase determine its function. *Science* **370**, doi:10.1126/science.abc2754 (2020).
4. Jumper, J. *et al.* Highly accurate protein structure prediction with AlphaFold. *Nature* **596**, 583-589, doi:10.1038/s41586-021-03819-2 (2021).
5. Senior, A. W. *et al.* Improved protein structure prediction using potentials from deep learning. *Nature* **577**, 706-710, doi:10.1038/s41586-019-1923-7 (2020).
6. Dror, R. O. *et al.* Structural basis for modulation of a G-protein-coupled receptor by allosteric drugs. *Nature* **503**, 295-299, doi:10.1038/nature12595 (2013).
7. Shan, Y. *et al.* Molecular basis for pseudokinase-dependent autoinhibition of JAK2 tyrosine kinase. *Nat Struct Mol Biol* **21**, 579-584, doi:10.1038/nsmb.2849 (2014).
8. Bussi, G. & Laio, A. Using metadynamics to explore complex free-energy landscapes. *Nat Rev Phys* **2**, 200-212, doi:10.1038/s42254-020-0153-0 (2020).
9. Henin, J., Lelievre, T., Shirts, M. R., Valsson, O. & Delemotte, L. Enhanced sampling methods for molecular dynamics simulations. *arXiv preprint arXiv:2202.04164* (2022).
10. Bolhuis, P. G., Chandler, D., Dellago, C. & Geissler, P. L. Transition path sampling: throwing ropes over rough mountain passes, in the dark. *Annu Rev Phys Chem* **53**, 291-318, doi:10.1146/annurev.physchem.53.082301.113146 (2002).
11. Schwartz, S. D. Perspective: Path Sampling Methods Applied to Enzymatic Catalysis. *J Chem Theory Comput* **18**, 6397-6406, doi:10.1021/acs.jctc.2c00734 (2022).
12. Laio, A. & Parrinello, M. Escaping free-energy minima. *Proc Natl Acad Sci U S A* **99**, 12562-12566, doi:10.1073/pnas.202427399 (2002).
13. Torrie, G. M. & Valleau, J. P. Non-Physical Sampling Distributions in Monte-Carlo Free-Energy Estimation - Umbrella Sampling. *J Comput Phys* **23**, 187-199, doi:Doi 10.1016/0021-9991(77)90121-8 (1977).
14. Darve, E. & Pohorille, A. Calculating free energies using average force. *Journal of Chemical Physics* **115**, 9169-9183, doi:Doi 10.1063/1.1410978 (2001).
15. Invernizzi, M., Piaggi, P. M. & Parrinello, M. Unified Approach to Enhanced Sampling. *Phys Rev X* **10**, doi:ARTN 041034
10.1103/PhysRevX.10.041034 (2020).
16. Levy, R. M., Srinivasan, A. R., Olson, W. K. & McCammon, J. A. Quasi-harmonic method for studying very low frequency modes in proteins. *Biopolymers* **23**, 1099-1112, doi:10.1002/bip.360230610 (1984).
17. Pietrucci, F., Marinelli, F., Carloni, P. & Laio, A. Substrate binding mechanism of HIV-1 protease from explicit-solvent atomistic simulations. *J Am Chem Soc* **131**, 11811-11818, doi:10.1021/ja903045y (2009).
18. Bonati, L., Piccini, G. & Parrinello, M. Deep learning the slow modes for rare events sampling. *Proc Natl Acad Sci U S A* **118**, doi:10.1073/pnas.2113533118 (2021).

19	Kang, P., Trizio, E. & Parrinello, M. Computing the committor with the committor to study the transition state ensemble. *Nat Comput Sci* **4**, 451-460, doi:10.1038/s43588-024-00645-0 (2024).
20	Bonati, L., Rizzi, V. & Parrinello, M. Data-Driven Collective Variables for Enhanced Sampling. *J Phys Chem Lett* **11**, 2998-3004, doi:10.1021/acs.jpclett.0c00535 (2020).
21	Belkacemi, Z., Gkeka, P., Lelievre, T. & Stoltz, G. Chasing Collective Variables Using Autoencoders and Biased Trajectories. *J Chem Theory Comput* **18**, 59-78, doi:10.1021/acs.jctc.1c00415 (2022).
22	Mardt, A., Pasquali, L., Wu, H. & Noe, F. VAMPnets for deep learning of molecular kinetics. *Nat Commun* **9**, 5, doi:10.1038/s41467-017-02388-1 (2018).
23	Kirmizialtin, S. & Elber, R. Revisiting and computing reaction coordinates with Directional Milestoning. *J Phys Chem A* **115**, 6137-6148, doi:10.1021/jp111093c (2011).
24	Pande, V. S., Beauchamp, K. & Bowman, G. R. Everything you wanted to know about Markov State Models but were afraid to ask. *Methods* **52**, 99-105, doi:10.1016/j.ymeth.2010.06.002 (2010).
25	Weinan, E., Ren, W. Q. & Vanden-Eijnden, E. String method for the study of rare events. *Phys Rev B* **66**, doi:ARTN 052301
10.1103/PhysRevB.66.052301 (2002).
26	Du, R., Pande, V. S., Grosberg, A. Y., Tanaka, T. & Shakhnovich, E. S. On the transition coordinate for protein folding. *J Chem Phys* **108**, 334-350, doi:Doi 10.1063/1.475393 (1998).
27	Ma, A. & Dinner, A. R. Automatic method for identifying reaction coordinates in complex systems. *J Phys Chem B* **109**, 6769-6779, doi:10.1021/jp045546c (2005).
28	Onsager, L. Initial recombination of ions. *Phys. Rev.* **54**, 554-557 (1938).
29	Ryter, D. On the Eigenfunctions of the Fokker-Planck Operator and of Its Adjoint. *Physica A* **142**, 103-121, doi:Doi 10.1016/0378-4371(87)90019-7 (1987).
30	Berne, B. J., Borkovec, M. & Straub, J. E. Classical and Modern Methods in Reaction-Rate Theory. *J Phys Chem-Us* **92**, 3711-3725, doi:DOI 10.1021/j100324a007 (1988).
31	Wu, S. & Ma, A. Mechanism for the rare fluctuation that powers protein conformational change. *J. Chem. Phys.* **156**, 05419 (2022).
32	Zheng, L. Q., Chen, M. G. & Yang, W. Random walk in orthogonal space to achieve efficient free-energy simulation of complex systems. *P Natl Acad Sci USA* **105**, 20227-20232, doi:10.1073/pnas.0810631106 (2008).
33	Zheng, L., Chen, M. & Yang, W. Simultaneous escaping of explicit and hidden free energy barriers: application of the orthogonal space random walk strategy in generalized ensemble based conformational sampling. *J Chem Phys* **130**, 234105, doi:10.1063/1.3153841 (2009).
34	Wu, S., Li, H. & Ma, A. Exact reaction coordinates for flap opening in HIV-1 protease. *PNAS* **119**, e2214906119, doi:https://doi.org/10.1073/pnas.2214906119 (2022).
35	Bolhuis, P. G., Dellago, C. & Chandler, D. Reaction coordinates of biomolecular isomerization. *Proc Natl Acad Sci U S A* **97**, 5877-5882, doi:10.1073/pnas.100127697 (2000).
36	Ma, A., Nag, A. & Dinner, A. R. Dynamic coupling between coordinates in a model for biomolecular isomerization. *J Chem Phys* **124**, 144911, doi:10.1063/1.2183768 (2006).

37  Li, W. & Ma, A. Recent developments in methods for identifying reaction coordinates. *Mol Simul* **40**, 784-793, doi:10.1080/08927022.2014.907898 (2014).

38  Best, R. B. & Hummer, G. Reaction coordinates and rates from transition paths. *P Natl Acad Sci USA* **102**, 6732-6737, doi:10.1073/pnas.0408098102 (2005).

39  Jung, H. *et al.* Machine-guided path sampling to discover mechanisms of molecular self-organization. *Nat Comput Sci* **3**, 334-345, doi:10.1038/s43588-023-00428-z (2023).

40  Li, W. & Ma, A. Reaction mechanism and reaction coordinates from the viewpoint of energy flow. *J Chem Phys* **144**, 114103, doi:10.1063/1.4943581 (2016).

41  Li, H. & Ma, A. Kinetic energy flows in activated dynamics of biomolecules. *J. Chem. Phys.* **153**, 094109, doi:doi: 10.1063/5.0020275 (2020).

42  Li, H., Wu, S. & Ma, A. Origin of protein quake: energy waves conducted by a precise mechanical machine. *J. Chem. Theory Comp.* **18**, 5692-5702 (2022).

43  Wu, S., Li, H. & Ma, A. A Rigorous Method for Identifying One-Dimensional Reaction Coordinate in Complex Molecules. *J. Chem. Theo. Comp.* **18**, 2836-2844 (2022).

44  Gianni, S. & Jemth, P. Allostery Frustrates the Experimentalist. *J Mol Biol* **435**, 167934, doi:10.1016/j.jmb.2022.167934 (2023).

45  Onsager, L. Reciprocal Relations in Irreversible Processes. II. *Phys. Rev.* **38**, 2265 - 2279 (1931).

46  Ishima, R., Freedberg, D. I., Wang, Y. X., Louis, J. M. & Torchia, D. A. Flap opening and dimer-interface flexibility in the free and inhibitor-bound HIV protease, and their implications for function. *Structure* **7**, 1047-1055, doi:10.1016/s0969-2126(99)80172-5 (1999).

47  Sadiq, S. K. & De Fabritiis, G. Explicit solvent dynamics and energetics of HIV-1 protease flap opening and closing. *Proteins* **78**, 2873-2885, doi:10.1002/prot.22806 (2010).

48  Miao, Y., Huang, Y. M., Walker, R. C., McCammon, J. A. & Chang, C. A. Ligand Binding Pathways and Conformational Transitions of the HIV Protease. *Biochemistry* **57**, 1533-1541, doi:10.1021/acs.biochem.7b01248 (2018).

49  Sadiq, S. K., Noe, F. & De Fabritiis, G. Kinetic characterization of the critical step in HIV-1 protease maturation. *Proc Natl Acad Sci U S A* **109**, 20449-20454, doi:10.1073/pnas.1210983109 (2012).

50  Dierynck, I. *et al.* Binding kinetics of darunavir to human immunodeficiency virus type 1 protease explain the potent antiviral activity and high genetic barrier. *J Virol* **81**, 13845-13851, doi:10.1128/JVI.01184-07 (2007).

51  Hummer, G. From transition paths to transition states and rate coefficients. *J Chem Phys* **120**, 516-523, doi:10.1063/1.1630572 (2004).

52  Jung, H., Okazaki, K. & Hummer, G. Transition path sampling of rare events by shooting from the top. *J Chem Phys* **147**, doi:Artn 152716
10.1063/1.4997378 (2017).

53  Barducci, A., Bussi, G. & Parrinello, M. Well-tempered metadynamics: a smoothly converging and tunable free-energy method. *Phys Rev Lett* **100**, 020603, doi:10.1103/PhysRevLett.100.020603 (2008).

54  Berezhkovskii, A. M. & Szabo, A. Diffusion along the splitting/commitment probability reaction coordinate. *J Phys Chem B* **117**, 13115-13119, doi:10.1021/jp403043a (2013).

55    Liu, X. & Fuentes, E. J. Emerging Themes in PDZ Domain Signaling: Structure, Function, and Inhibition. *Int Rev Cell Mol Biol* **343**, 129-218, doi:10.1016/bs.ircmb.2018.05.013 (2019).

56    Peterson, F. C., Penkert, R. R., Volkman, B. F. & Prehoda, K. E. Cdc42 regulates the Par-6 PDZ domain through an allosteric CRIB-PDZ transition. *Mol Cell* **13**, 665-676, doi:10.1016/s1097-2765(04)00086-3 (2004).

57    Bozovic, O., Jankovic, B. & Hamm, P. Sensing the allosteric force. *Nat Commun* **11**, 5841, doi:10.1038/s41467-020-19689-7 (2020).

58    Petit, C. M., Zhang, J., Sapienza, P. J., Fuentes, E. J. & Lee, A. L. Hidden dynamic allostery in a PDZ domain. *Proc Natl Acad Sci U S A* **106**, 18249-18254, doi:10.1073/pnas.0904492106 (2009).

59    Bozovic, O. *et al.* Real-time observation of ligand-induced allosteric transitions in a PDZ domain. *Proc Natl Acad Sci U S A* **117**, 26031-26039, doi:10.1073/pnas.2012999117 (2020).

60    Buchli, B. *et al.* Kinetic response of a photoperturbed allosteric protein. *Proc Natl Acad Sci U S A* **110**, 11725-11730, doi:10.1073/pnas.1306323110 (2013).

61    Huang, J. *et al.* CHARMM36m: an improved force field for folded and intrinsically disordered proteins. *Nat Methods* **14**, 71-73, doi:10.1038/nmeth.4067 (2017).

62    Jorgensen, W. L., Chandrasekhar, J., Madura, J. D., Impey, R. W. & Klein, M. L. Comparison of Simple Potential Functions for Simulating Liquid Water. *J Chem Phys* **79**, 926-935, doi:Doi 10.1063/1.445869 (1983).

63    Hess, B., Kutzner, C., van der Spoel, D. & Lindahl, E. GROMACS 4: Algorithms for highly efficient, load-balanced, and scalable molecular simulation. *J Chem Theory Comput* **4**, 435-447, doi:10.1021/ct700301q (2008).

64    Hess, B., Bekker, H., Berendsen, H. J. C. & Fraaije, J. G. E. M. LINCS: A linear constraint solver for molecular simulations. *J Comput Chem* **18**, 1463-1472, doi:Doi 10.1002/(Sici)1096-987x(199709)18:12<1463::Aid-Jcc4>3.0.Co;2-H (1997).

65    Tiwary, P. & Berne, B. J. Spectral gap optimization of order parameters for sampling complex molecular systems. *Proc Natl Acad Sci U S A* **113**, 2839-2844, doi:10.1073/pnas.1600917113 (2016).

66    Tribello, G. A., Bonomi, M., Branduardi, D., Camilloni, C. & Bussi, G. PLUMED 2: New feathers for an old bird. *Comput Phys Commun* **185**, 604-613, doi:10.1016/j.cpc.2013.09.018 (2014).

67    Pietrucci, F., Marinelli, F., Carloni, P. & Laio, A. Substrate Binding Mechanism of HIV-1 Protease from Explicit-Solvent Atomistic Simulations. *J Am Chem Soc* **131**, 11811-11818, doi:10.1021/ja903045y (2009).

**Enhanced Sampling of Protein Conformational Changes via True Reaction Coordinates from Energy Relaxation**

Huiyu Li and Ao Ma*

**Supplementary Information**

Methods

Fig. S1, S2, S3, S4, S5, S6

Supplementary Videos S1 to S5

**Clustering trajectories into ensembles.** To divide the 8,000 trajectories into ensembles with k-means clustering, we first normalize all the PEF trajectories as: $\Delta \widetilde{W}_i(0 \to n\delta t; \alpha) = \frac{\Delta W_i(0 \to n\delta t; \alpha)}{|\Delta W_i^{max}(0 \to t; \alpha)|}$, where $\Delta W_i^{max}(0 \to t; \alpha)$ is the highest magnitude of $\Delta W_i$ along trajectory $\alpha$. In this way, we have $\Delta \widetilde{W}_i(0 \to n\delta t; \alpha) \in [-1, 1]$. We then decide the number of clusters by visual inspection of PEF of $q_i$ along different trajectories. Six seed PEF trajectories that represent distinct patterns were selected. To determine clusters, we define the distance between two trajectories $\alpha$ and $\beta$ as: $s_{\alpha\beta} = \frac{1}{N}\sum_{n=1}^{N}[\Delta \widetilde{W}_i(0 \to n\delta t; \alpha)\Delta \widetilde{W}_i(0 \to n\delta t; \beta)]$, where $N = 5{,}000$ is the number of time steps in a trajectory.

The clustering procedure involves the following steps:

1) Compute the distance between each trajectory α and each of the six seed trajectories. Assign α to cluster-*i* if its distance to seed trajectory *i* is the shortest. This procedure divides the 8,000 trajectories into six initial clusters.

2) Re-evaluate each trajectory's distance to each of the initial clusters obtained in step one. The distance between trajectory α and cluster-*i* is calculated as the average distance between α and all the trajectories within cluster-*i*. Re-assign α to the cluster it is closest to.

3) Repeat step two until there are no changes in the cluster assignments.

This clustering procedure was applied to all internal coordinates and singular coordinates. With this procedure, the 8,000 trajectories are divided into six ensembles, $E_0(q_i), \ldots, E_5(q_i)$. Among them, $E_0(q_i)$ and $E_1(q_i)$ contain ~3,300 trajectories, $E_2(q_i)$ and $E_3(q_i)$ contain ~550 trajectories, $E_4(q_i)$ and $E_5(q_i)$ contain ~150 trajectories, respectively. Examples of potential energy flows of a backbone dihedral along trajectories in these ensembles are shown in Fig. S1. The GWF analysis was conducted for trajectories in $E_0(q_i)$ and $E_1(q_i)$, as they contain more trajectories. The results are identical.

**One-dimensional FEP along a single tRC.** An important proof-of-concept is to calculate the FEP along individual tRCs. However, our method for applying bias potential does not support unweighting with existing schemes. Therefore, we used umbrella sampling to calculate the FEP along $u_0$. Specifically, we divided $\Delta u_0$, the change in $u_0$ relative to the crystal structure, from 0 to 2.6 radians into 14 intervals. We selected protease conformations with proper $\Delta u_0$ values from the RC-uncovered trajectory segment that corresponds to the first flap opening process in Fig. 5 above.

For each umbrella window, we apply a harmonic potential to restrain $u_0$ and run a 1 ns MD trajectory, sampling one conformation every 100 fs. We also tested sampling every 500 fs, and obtained the same FEP result. We then used weighted histogram analysis to generate the FEP, shown in Fig. S2. The red line represents the FEP, and the blue lines indicate the standard deviation computed using bootstrapping.

The FEP along $u_0$ displays a clear double-well feature. Aside from the flap-closed state, there is a metastable state with a minimum at $\Delta u_0 = 2.2$ radian, corresponding to flap distance of 2.2 to 2.5 nm based on the trajectory in Fig. 5. Because the underlying FES is high dimensional, using this 1D FEP along $u_0$ to interpret the mechanism of flap opening and ligand dissociation requires caution. Nevertheless, we found the insights suggested by this 1D FEP to be physically sensible, as outlined below.

The initial phase of flap-opening and ligand dissociation shows three distinct features: rapid increase in $u_0$, lack of significant change in flap distance $d_f$, and dislodging of the ligand from the active site (Figs. 4, 5, 6, S3). This phase corresponds to climbing the energy barrier on the 1D FEP. Due to the lack of flap distance change, there is no large-scale structural rearrangement. Therefore, this process does not involve water reorganization, making protein-protein interactions the dominating force. The bias force applied to $u_0$ overcomes resistance from non-RCs, effectively accelerating the motion of $u_0$. Due to the collective and cooperative nature of $u_0$, the ligand's motion is tightly coupled to the protein backbone, resulting in its dislodgement from the active site.

The next phase features slowdown in the increase of $u_0$ and a rapid rise in $d_f$, accompanied by further ligand movements (Fig. 5, 6, S3). This phase corresponds to descending from the barrier top into the basin of the flap-open state on the 1D FEP, where the lowest free energy state is located at $\Delta u_0 = 2.2$ radian, corresponding to $d_f = 2.5$ nm based on Fig. 5. The large-scale flap movement requires significant water reorganization, generating strong resistance from water molecules. Since the bias force only acts on the protein, it is less effective at overcoming water resistance, leading to the observed slowdown in $u_0$. As flaps open, water molecules begin to flux into the active site, driving additional ligand movements.

The third phase is specific to ligand dissociation, characterized by fluctuations in $d_f$ and dissociation of the ligand (Fig. 5, S3). The $d_f$ fluctuation corresponds to movements within the basin of the flap-open state and are driven by water fluxing into the active site to solvate the ligand. Along RC-uncovered trajectories, $d_f$ fluctuations are small, as the biasing force suppresses fluctuations in $u_0$ and keeps $d_f$ at a large value. The extent of flap opening during this phase depends on the ligand size—larger ligand requires greater flap opening, with $d_f = 2.5$ nm for DRV and 3 nm for MA/CA. In contrast, NRTs exhibit much more pronounced $d_f$ fluctuations, primarily drive by water motion. Both the value of $d_f$ and the extent of its fluctuation depend on ligand size: larger ligand requires larger-scale water movement for solvation, which leads to larger impact on protein structure, reflected in larger $d_f$ fluctuations.

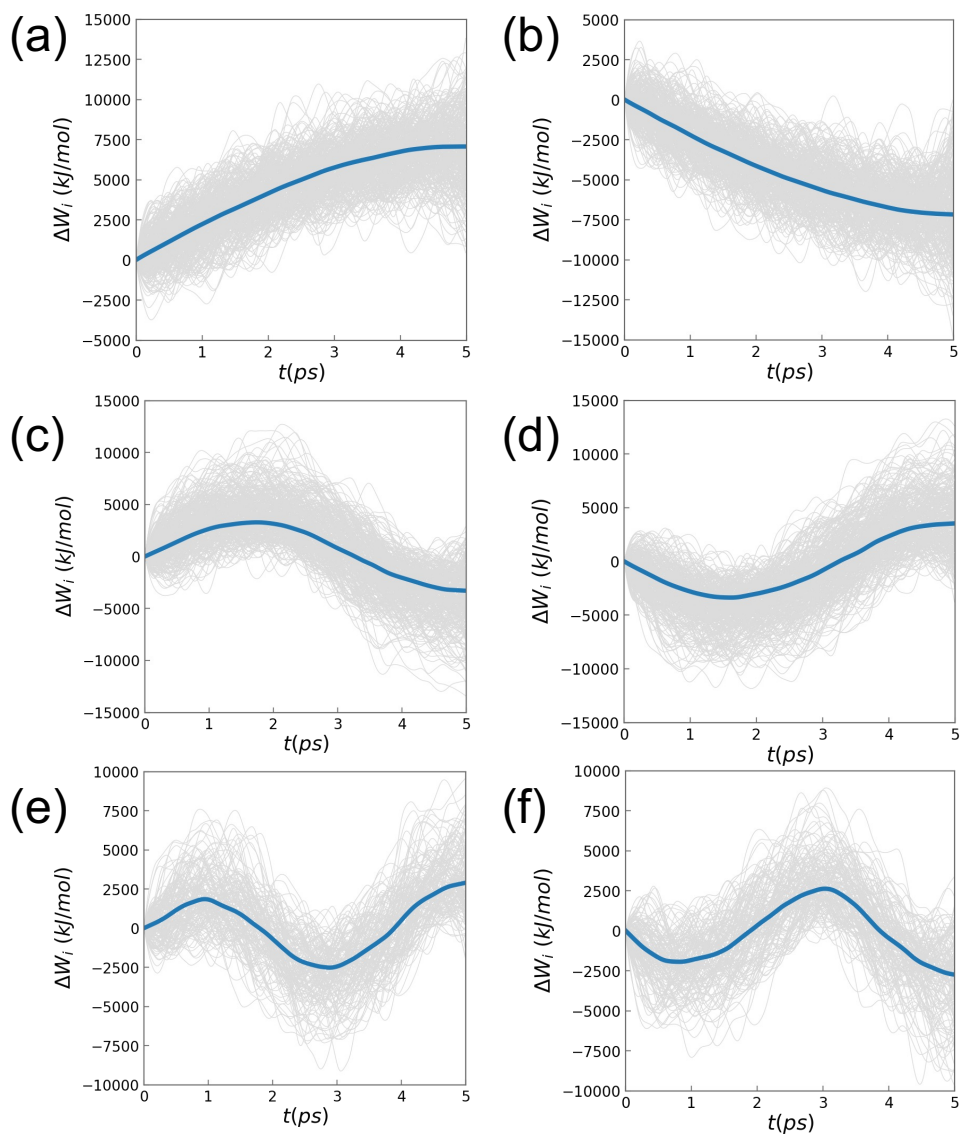

**Fig. S1: Sample trajectories of the potential energy flow of a backbone dihedral from different ensembles.** The gray lines are potential energy flows along individual trajectories. The thick blue lines are the averaged potential energy flows of $E_0(q_i)$ **(a)**, $E_1(q_i)$ **(b)**, $E_2(q_i)$ **(c)**, $E_3(q_i)$ **(d)**, $E_4(q_i)$ **(e)** and $E_5(q_i)$ **(f)**, respectively.

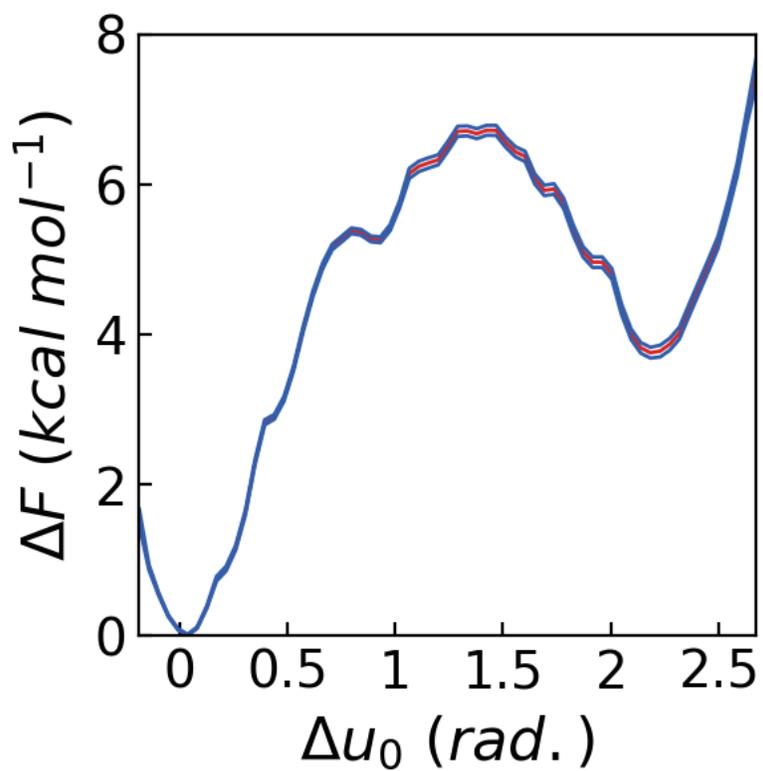

**Figure S2**: One-dimensional free energy profile along $u_0$ calculated using umbrella sampling. The red line is the computed free energy profile and the blue lines are the standard deviation estimated using bootstrapping.

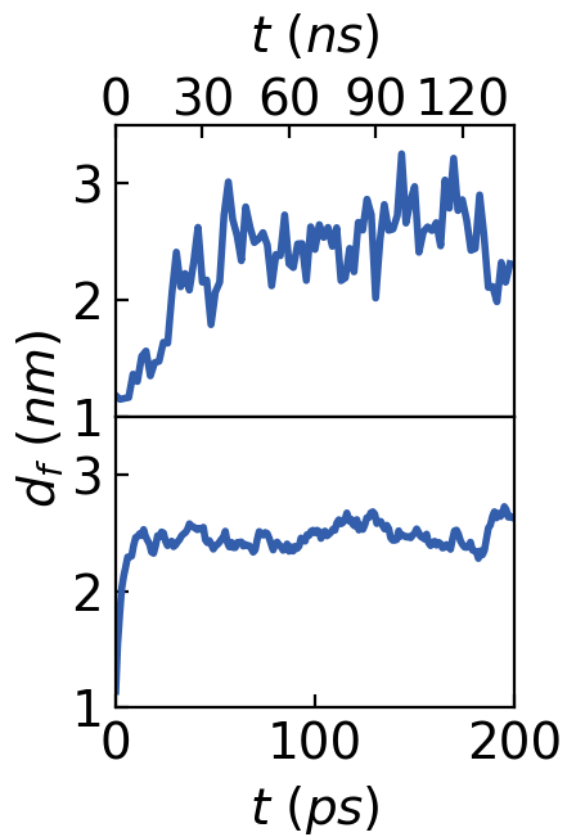

**Figure S3**: Time evolution of the flap distance $d_f$ along the NRT (upper panel) and RC-uncovered trajectory (lower panel) of DRV dissociation from HIV-PR.

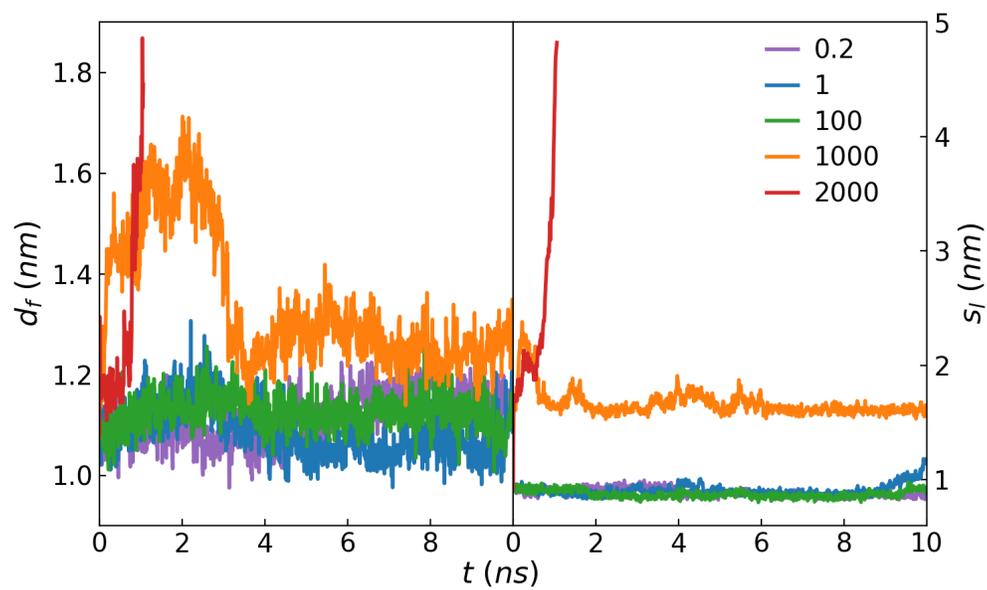

**Figure S4**: Time evolution of $d_f$ (left) and $s_l$ (right) along metadynamics trajectories obtained with Gaussian heights of 0.2, 1, 100, 1,000 and 2,000 kJ mol$^{-1}$.

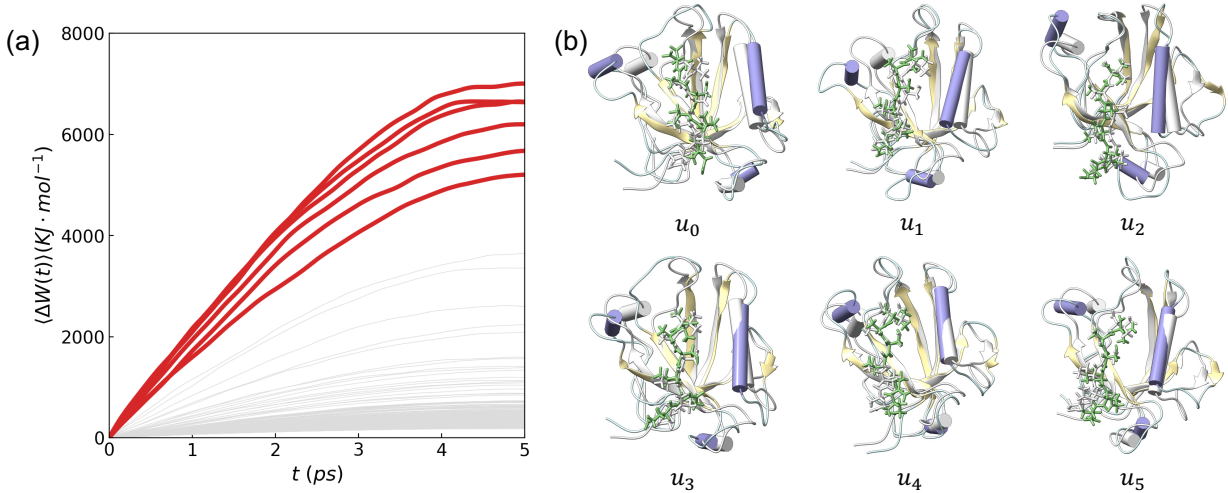

**Fig. S5**: **(a)** Potential energy flows of SCs of PDZ2 energy relaxation. **(b)** End structures on RC-uncovered trajectories along the six RCs for PDZ2. The crystal structure of PDZ2 is shown in White color as comparison to highlight the conformational changes and the dislocation of the ligand. Ligand in conformations from RC-uncovered trajectories are shown in Green.

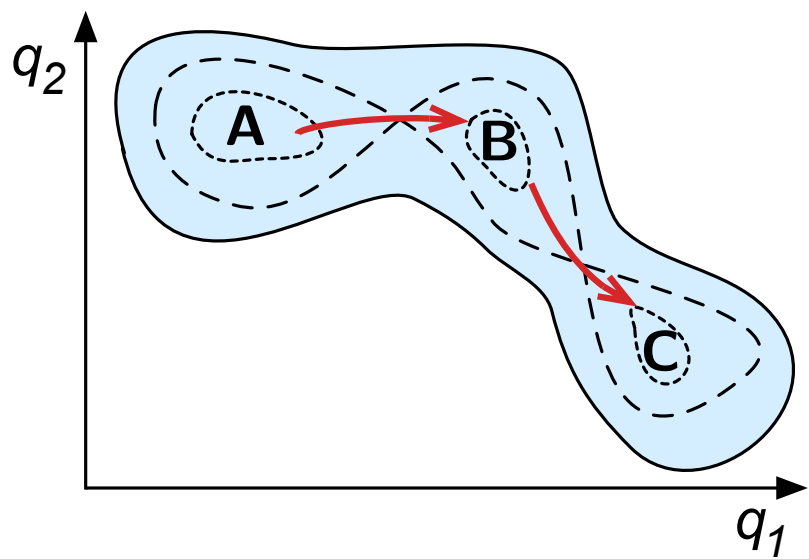

**Fig. S6**: Procedure to apply of our method to a multi-step reaction. Our method can be applied to a multi-step process by following the standard protocol of chemical kinetics, using the two-step process shown above as an example. Assuming the barriers for both $A \rightarrow B$ and $B \rightarrow C$ transitions are much higher than thermal energy, energy relaxation is well-defined in both basins A and B. Starting with the known conformation A, we will first simulate energy relaxation in basin A to compute the tRC for the $A \rightarrow B$ transition (red arrow pointing from A to B). We can then generate RC-uncovered and natural reactive trajectories for the $A \rightarrow B$ transition, allowing us to determine the equilibrium structures in basin B. Next, we will simulate energy relaxation in basin B and compute the tRC for $B \rightarrow C$ transition (red arrow pointing from B to C), enabling us to simulate this transition. In this way, we can generate ensembles of RC-uncovered and NRT trajectories for both $A \rightarrow B$ and $B \rightarrow C$ transitions, which can be used to compute thermodynamic and kinetic variables for both transitions. Afterwards, the overall rate for $A \rightarrow B \rightarrow C$ can be obtained by solving the corresponding reaction rate equations following the standard procedures of chemical kinetics.

**Supplementary Videos**

1. Supplementary video S1: movie of an RC-uncovered trajectory of HIV-PR bound to DRV obtained by applying a bias potential along $u_0^D$ with a force constant of $k = 30,000$ kJ/mol. The duration of the trajectory is 200 ps. Snapshots of this trajectory are shown in Fig. 4a. As shown by the time evolution of the flap distance $d_f$ in Fig. S3 (lower panel), the process begins with a rapid opening of the flaps to $d_f = 2.5$ nm, accompanied by ligand dislodging from the active site. Afterwards, the ligand undergoes diffusive motion and the flaps fluctuate in the wide-open state, both driven by the influx of water molecules into the active site to solvate the ligand. Ligand dissociation completes when it is solvated and exits the active site. For visualization purposes, water molecules are not shown.

2. Supplementary video S2: movie of an RC-uncovered trajectory of HIV-PR bound to MA/CA peptide obtained by applying a bias potential along $u_0^M$ with a force constant of $k = 30,000$ kJ/mol. The duration of the trajectory is 200 ps. Snapshots of this trajectory are shown in Fig. 4c. As shown by the time evolution of the flap distance $d_f$ in Fig. 6c, the process begins with a rapid opening of the flaps to $d_f = 3$ nm, accompanied by ligand dislodging from the active site. Afterwards, the ligand undergoes diffusive motion and the flaps fluctuate in the wide-open state, both driven the influx of water molecules into the active site to solvate the ligand. Ligand dissociation completes when it is solvated and exits the active site. For visualization purposes, water molecules are not shown.

3. Supplementary video S3: movie of an NRT of DRV dissociation from HIV-PR. The duration of the trajectory is 140 ns. Snapshots of this trajectory are presented in Fig. 4b. As shown by the time evolution of the flap distance $d_f$ in Fig. S3, the process begins with a gradual opening of the flaps to $d_f = 2.5$ nm, accompanied by ligand dislodging from the active site. Afterwards, the ligand undergoes diffusive motion and $d_f$ fluctuates between 2.2 and 2.8 nm, both driven the influx of water molecules into the active site to solvate the ligand. Eventually, ligand dissociation completes when it becomes fully solvated and diffuses into bulk water. For visualization purposes, water molecules are not shown.

4. Supplementary video S4: movie of an NRT of MA/CA peptide dissociation from HIV-PR. The duration of the trajectory is 540 ns. Snapshots of this trajectory are presented in Fig. 4d. As shown by the time evolution of the flap distance $d_f$ in Fig. 6c, the process begins with a period of thermal fluctuations in the flap-closed state, followed by gradual opening of the flaps to $d_f = 2.5$ nm and ligand dislodging from the active site. Afterwards, the ligand undergoes diffusive motion and $d_f$ fluctuates widely between 2.0 and 3.2 nm due to the large size of the MA/CA peptide, both driven the influx of water molecules into the active site to solvate the ligand. Eventually, ligand dissociation completes when it becomes fully solvated and diffuses into bulk water. For visualization purposes, water molecules are not shown.

5. Supplementary video S5: movie of an NRT of a 6-residue (EQVSAV) peptide dissociation from PDZ2. The duration of the trajectory is 48 ns. Snapshots of this trajectory are presented in Fig. 8. As depicted by the time evolution of $d_1$ and $d_2$ in Fig. 9a, the process begins with fluctuations in both the Cdc42 binding interface and the ligand binding groove. This is followed by a rapid increase in the distance between $\alpha_1$ helix and $\beta_1$ sheet, reaching $d_1 = 12$ Å where it fluctuates, triggering the ligand to dislodge from the binding groove.

At the transition state, the binding groove starts to widen due to the outward motion of $\alpha_2$ helix (see Fig. 8) and $d_1$ starts to drop. When $d_2$ reaches 16 Å and $d_1$ drops back to its crystal structure value, the ligand starts to dissociate from the binding groove. After complete dissociation, the $\alpha_2$ helix snaps back, returning the protein to its crystal structure conformation.